\documentstyle{mn}
\newif\ifAMStwofonts  
  
\newcommand{\Msolar}{M$_{\odot}$}                           
\newcommand{\Myear}{M$_{\odot}$ yr$^{-1}$}                

\ifoldfss  
  \ifCUPmtlplainloaded \else  
    \NewTextAlphabet{textbfit} {cmbxti10} {}  
    \NewTextAlphabet{textbfss} {cmssbx10} {}  
    \NewMathAlphabet{mathbfit} {cmbxti10} {} 
    \NewMathAlphabet{mathbfss} {cmssbx10} {} 
  \fi  
  \ifAMStwofonts  
    \ifCUPmtlplainloaded \else  
      \NewSymbolFont{upmath} {eurm10}  
      \NewSymbolFont{AMSa} {msam10}  
      \NewMathSymbol{\upi}     {0}{upmath}{19}  
      \NewMathSymbol{\umu}     {0}{upmath}{16}  
      \NewMathSymbol{\upartial}{0}{upmath}{40}  
      \NewMathSymbol{\leqslant}{3}{AMSa}{36}  
      \NewMathSymbol{\geqslant}{3}{AMSa}{3E}

       \let\le=\leqslant  
         
    \fi  
  \fi  
\fi 
  
\ifnfssone  
  \newmathalphabet{\mathit}  
  \addtoversion{normal}{\mathit}{cmr}{m}{it}  
  \addtoversion{bold}{\mathit}{cmr}{bx}{it}  
  \newmathalphabet{\mathbfit} 
  \addtoversion{normal}{\mathbfit}{cmr}{bx}{it}  
  \addtoversion{bold}{\mathbfit}{cmr}{bx}{it}  
  \newmathalphabet{\mathbfss} 
  \addtoversion{normal}{\mathbfss}{cmss}{bx}{n}  
  \addtoversion{bold}{\mathbfss}{cmss}{bx}{n}  
  \ifAMStwofonts  
    \ifCUPmtlplainloaded \else  
      \UseAMStwoboldmath  
      \makeatletter  
      \new@mathgroup\upmath@group  
      \define@mathgroup\mv@normal\upmath@group{eur}{m}{n}  
      \define@mathgroup\mv@bold\upmath@group{eur}{b}{n}  
      \edef\UPM{\hexnumber\upmath@group}  
      \new@mathgroup\amsa@group  
      \define@mathgroup\mv@normal\amsa@group{msa}{m}{n}  
      \define@mathgroup\mv@bold\amsa@group{msa}{m}{n}  
      \edef\AMSa{\hexnumber\amsa@group}  
      \makeatother  
      \mathchardef\upi="0\UPM19  
      \mathchardef\umu="0\UPM16  
      \mathchardef\upartial="0\UPM40  
      \mathchardef\leqslant="3\AMSa36  
      \mathchardef\geqslant="3\AMSa3E  

       \let\le=\leqslant  

    \fi  
  \fi  
\fi 
  
\ifnfsstwo  
  \DeclareMathAlphabet{\mathbfit}{OT1}{cmr}{bx}{it}  
  \SetMathAlphabet\mathbfit{bold}{OT1}{cmr}{bx}{it}  
  \DeclareMathAlphabet{\mathbfss}{OT1}{cmss}{bx}{n}  
  \SetMathAlphabet\mathbfss{bold}{OT1}{cmss}{bx}{n}  
  \ifAMStwofonts  
    \ifCUPmtlplainloaded \else  
      \DeclareSymbolFont{UPM}{U}{eur}{m}{n}  
      \SetSymbolFont{UPM}{bold}{U}{eur}{b}{n}  
      \DeclareSymbolFont{AMSa}{U}{msa}{m}{n}  
      \DeclareMathSymbol{\upi}{0}{UPM}{"19}  
      \DeclareMathSymbol{\umu}{0}{UPM}{"16}  
      \DeclareMathSymbol{\upartial}{0}{UPM}{"40}  
      \DeclareMathSymbol{\leqslant}{3}{AMSa}{"36}  
      \DeclareMathSymbol{\geqslant}{3}{AMSa}{\"3E}  

       \let\le=\leqslant  

    \fi  
  \fi  
\fi 
  
\ifCUPmtlplainloaded \else  
  \ifAMStwofonts \else 
    \def\upi{\pi}  
    \def\umu{\mu}  
    \def\upartial{\partial}  
  \fi  
\fi

\title[Multiwavelength Study of the Chandra Deep Field South
]
{Multiwavelength Study of X-ray Selected Star 
Forming Galaxies within the  Chandra Deep Field South
}  
  
\author[Rosa--Gonz\'alez et al.]  
{
Daniel Rosa--Gonz\'alez$^{1}$, Denis Burgarella$^{2}$, Kirpal Nandra$^{3}$, 
Daniel Kunth$^{4}$,   
 \\ \\ {\LARGE Elena Terlevich$^{1}$\thanks{Research Affiliate IoA, Cambridge, UK}  and 
Roberto Terlevich$^{1}$\thanks{Research Affiliate IoA, Cambridge, UK}}  \\
$^1$ INAOE, Luis Enrique Erro 1. Tonantzintla, Puebla 72840. M\'exico.\\
$^2$Observatoire Astronomique Marseille Provence, Laboratoire d'Astrophysique de Marseille, 
13376 Marseille Cedex 12, France\\
$^3$Astrophysics Group, Imperial College London, Blackett Laboratory, Prince Consort Road, London SW7 2AW\\
$^4$Institut d'Astrophysique, Paris, 98 bis Boulevard Arago, F-75014 Paris, France\\
}

\date{Accepted  .  
      Received ;  
      in original form \today\  (CDFS-vs202)}  
  
\pagerange{\pageref{firstpage}--\pageref{lastpage}}  
\pubyear{2004}  
  
\begin{document}  
  
\maketitle  
  
\label{firstpage}  
  
\begin{abstract}  
We have combined multi-wavelength observations of a selected sample of
starforming galaxies with galaxy evolution models in order to compare the
results obtained for different
SFR tracers and to study the effect that the evolution 
of the starforming regions has on them. 
We also aimed at obtaining a better understanding of the corrections due to
extinction and nuclear activity on the derivation of the SFR.
We selected the sample from Chandra data for the well studied region 
Chandra Deep Field South
(CDFS) and chose the objects that also have  UV and IR data from 
GALEX and GOODS-Spitzer respectively. 

Our main finding is that there is good agreement between the extinction
corrected SFR(UV) and the  SFR(X), and
we confirm the use of X-ray luminosities as a trustful
tracer of recent star formation activity.
Nevertheless, at SFR(UV) larger than about 5\Myear\ there are
several galaxies with an excess of SFR(X) 
suggesting  the presence of an obscured AGN not detected in the 
optical spectra. We conclude that the IR luminosity is driven
by recent star formation even in those galaxies  where the
SFR(X) is an order of magnitude higher than the SFR(UV) and therefore
may harbour an AGN. One object shows SFR(X) much lower than 
expected based on the SFR(UV); this
SFR(X) `deficit'  may be due to an early transient phase
before most of the massive X-ray binaries were formed.
An X-ray deficit could be used to select
extremely young bursts in an early phase just after the explosion of
the first supernovae associated with massive stars and before the onset of
massive X-ray binaries.
\end{abstract}  
  
\section{Introduction}  

The UV emission in galaxies traces the presence of massive stars  
related to a recent episode of star formation. For this 
reason,  the UV luminosity is widely used as a tracer of the star formation 
activity not only in nearby galaxies but also in the early universe
(e.g. Lilly et al. 1995, Madau et al. 1996).  
However it is important to bear in mind that the corrections applied to the 
observed UV fluxes (e.g.~due to extinction and evolutionary 
uncertainties) are neither unique nor 
straightforward implying that multiwavelength studies are 
necessary to improve the correction factors 
(e.g. Buat et al. 2005, Schmitt et al. 2006).
 
The X-ray emission in normal galaxies -- without an AGN component -- 
is produced by  high- and low-  mass X-ray binaries, young supernova
remnants  and cooling hot gas (e.g. Fabbiano 1989).
The relation between the observed X-ray luminosities and 
the current SFR is driven by the presence of 
High Mass X-ray Binaries (HMXB) where the primary star 
is a collapsed object with mass higher than 2.5 \Msolar\ 
and the secondary star is a massive star classified as O, B or Wolf Rayet
(Van~Bever and Vanbeveren 2000, Cervi{\~ n}o, Mas-Hesse and Kunth 2002).
Due to the strong correlation found between  the X-ray luminosity and 
other tracers of recent star formation activity such as the 
FIR luminosity (David, Jones and Forman  1992) the 
radio emission (Ranalli, Comastri and Setti 2003) and the number of  
HMXB (Grimm et al. 2003), the X-ray luminosity
 has been used to calculate the star formation 
rate (SFR) even at high redshift  (Nandra et al. 2002, Laird et al. 2005). 
However at large distances, it is difficult to separate the different 
contributors to the X-ray luminosity,
hence the estimated SFR could be jeopardized by 
the presence of obscured  
AGN unnoticed in the optical spectral range and also 
by low  mass X-ray binaries (LMXB) which 
are not related to the recent star formation event.

In addition, in very young systems we expect to observe a deviation of the 
SFR given by the X-ray luminosities [SFR(X)] 
with respect to the SFR given by the UV [SFR(UV)] 
due to the different time scales for the emission from massive 
stars -- responsible for the UV (e.g.~Mas-Hesse and Kunth 1991) -- 
and the formation of the first HMXB which dominate the X-ray 
luminosities.
These different time scales would produce a time lag between the 
UV and X-ray emission -- depending on the upper mass of the initial mass 
function of the ionizing cluster --  of at least a few million years.
Notice that stars with masses higher than about 8\Msolar\  end their lives
as supernovae after about 40 Myr when they might form a binary system. 
Therefore, the existence of a time lag  between the formation of the
massive stellar cluster and the formation of the first 
compact object that could end as a binary system  
opens an oportunity to use the lack of X-rays  to search for 
young objects of between 4 and up to $\sim$40 Myr. 
The lag between the formation of the first massive stars and the formation of 
compact objects has been already successfully used to find young objects using 
deep radio observations in which -- as in the case of X-rays -- 
the radio emission from supernova remnants is produced after some million years
of the formation of the stellar cluster (Rosa-Gonz\'alez et al. 2007).

Multi-wavelength observations of the  Chandra Deep Field South (CDFS) -- 
one of the best studied patches of the sky --
have contributed to our understanding of fundamental processes in
galaxy evolution (e.g. Giacconi et al. 2001, Tozzi et al. 2001, 
Gabasch et al. 2004,  Adami et al. 2005, Ferreras et al. 2005). 
In this paper we  focus on the relation between 
the observed UV and X-rays based on archival data provided by the 
{\it Great Observatories Origins Deep Survey} (GOODS)
for the CDFS, to study in detail the relation between the SFR(UV)
and the SFR(X).
We also include data obtained from the Spitzer archives 
for estimating the star formation based on  Infrared luminosities [SFR(IR)]
and to put stringent limits on the extinction corrections 
applied to the UV fluxes.

This paper is organized in the following way: 
in Section~\ref{Sec:Sample} we describe the galaxy sample; the method 
adopted to obtain 
UV  fluxes is presented in Section~\ref{Uvfluxes}, and in
Section~\ref{Sec:uvX} we obtain the star formation rate
from UV and X-ray luminosities. The contamination by AGN and the 
existence of fainter than expected X-ray galaxies are discussed in 
Section~\ref{AGN}.
Section~\ref{Spitzer} describes the FIR data of our subsample of galaxies
detected with Spitzer and the derived SFR(IR).
Conclusions are presented in Section~\ref{Sec:Conclusions}.

Throughout this work a standard, flat $\Lambda$CDM cosmology with
$\Omega_\Lambda$= 0.7 and H$_0$ = 70 km s$^{-1}$ Mpc$^{-1}$ is assumed.

\section{The sample}\label{Sec:Sample}

The Chandra Deep Field South (CDFS), centered in coordinates 
$\alpha_{2000}=03^h32^m25^s$ and 
$\delta_{2000}=$-27$^{\rm o}$48\arcmin50\arcsec,
was observed during 942 ksec using the Advanced CCD Imaging Spectrometer on 
board the Chandra X-ray observatory.
The CDFS data also  include optical identifications based on 
{\it R} band imaging of the field 
(Giacconi et al. 2002, Rosati et al. 2002, Alexander et al. 2003). 
We cross--correlated the Chandra observations of the CDFS  
with UV data from the {\it Galaxy Evolution Explorer} 
(GALEX). 

GALEX is a NASA Small Explorer Mission launched on the 28th.~of April 2003
developed in cooperation with the Centre National d'Etudes
Spatiales of France and the South Korean Ministry of Science and
Technology.
It was 
designed to perform several surveys in two ultraviolet bands: FUV
centered at 1530\AA\ and NUV at 2310\AA. The detailed
characteristics of GALEX are given in Morrissey et al. (2005). 
In brief, with a very wide $1.^{\rm o}25$ field of view and an angular
resolution of $\sim$ 5\arcsec (FWHM), GALEX is performing the first
ultraviolet All-sky Imaging Survey (AIS) down to $m_{AB}=20.5$. 
The GALEX data  at 2312~\AA\, (Near-UV), and 1522~\AA\, (Far-UV) 
used in this work were obtained directly from the 
MAST archive\footnote{Multimission archive at STScI;
http://archive.stsci.edu/} and they are part of the 
GALEX Deep Imaging Survey project which reaches a magnitude limit of 
$m_{AB}=25$.

Alexander et al. (2003) listed 326 CDFS sources of which 293 were detected in 
the soft band (0.5--2 kev), 198 in the hard band (2--10 kev) and 174 in both.
We searched for the GALEX counterpart in a circle of 5\arcsec\ 
radius including the position errors both in  GALEX and in Chandra. 
We were left with a preliminary subsample of 59 objects that includes only 
those detected in the two GALEX bands.
Some objects that are just upper limits in one or both of Chandra bands were 
still retained in the analysis that follows. 
     Notice that  these weak sources were extracted from the main
     catalog  by Alexander et al. (2003) where an X-ray source is
     defined as an object that is detected with a false-positive
     probability higher than 10$^{-7}$, 
     at least in one of the seven standard bands defined in their paper.
This first selection, would have 
excluded most of the obscured AGNs contained in early type galaxies, and also 
starforming galaxies with redshift larger than about 1. 
The following step was to use the  broad photometric
classification  given by the COMBO-17 
team \footnote{COMBO-17 (Classifying Objects by Medium-Band Observations
- a spectrophotometric 17-filter survey) is a photometric survey perfomed with 
the MPG/ESO 2.2-m telescope at La Silla, Chile (Wolf et al.~2004).}
to select only  objects classified as galaxies 
excluding QSO or Seyfert~1. 
The COMBO-17 classification is based on both the size of the image, 
and the  colours of the object. 
Objects with clearly extended morphology are classified as galaxies while
compact objects which present a power law continuum are 
classified as QSO (Wolf et al.~2004). After inspection of the provided 
images (optical and UV) we removed those objects for which the GALEX 
source was not univocally associated to an optical counterpart.
In table~\ref{Tab:CDFS-X-uv} we present the final sample of 29 galaxies
including CDFS132, for which both X-ray fluxes are considered
upper limits.

The separations between 
the Chandra and the corresponding GALEX source are smaller than 3\arcsec\
except for  CDFS213 and  CDFS325. For these galaxies we check
the positions given by  Giacconi et al. (2002) and found that 
they lie closer to the optical couterpart. 
Different position determinations are expected when different extraction 
methods are used on difuse weak sources. 
The  fluxes provided by these authors for these sources are similar 
in both soft and hard band;  for the 
soft X-rays flux of CDFS213, Alexander et al. considered the value 
as an upper limit while Giacconi et al. give the soft X-ray flux as a 
detection.

As we mentioned above, some of the Chandra 
detections are only upper limits but we left them in 
the sample as they have been detected by GALEX in the two UV bands.

Table~\ref{Tab:CDFS-X-uv} lists UV and X-ray fluxes of the selected objects.
It also includes the corresponding {\it R} band and the photometric redshift  
based on COMBO-17 observations when spectroscopic data was not avaliable. 
The COMBO-17 observations of the CDFS include five different broad band
filters (U,B,V,R and I) and 12 narrow filters necessary to constrain the 
photometric redshifts with an accuracy of about 5\% at z$\la$1 
(Wolf et al.~2004).
The optical broad band data are presented in Table~\ref{Tab:CDFS-Optical}.

The spectroscopic redshift (z$_{sp}$) comes from Szokoly et al. (2004) 
except for 
CDFS132 and CDFS185 for which z$_{sp}$ comes from the recent work by 
Ravikumar et al. (2006).

\begin{table*}
\begin{center}\caption{\label{Tab:CDFS-X-uv}
X-ray and UV characteristics of the selected objects. The columns are: object 
name, separation 
between Chandra  and GALEX positions (Chandra positions are from  Alexander 
et al. 2003; in parenthesis, distances using positions by Giacconi 
et al. (2002) ), flux and luminosity in the soft and hard X-ray bands$^1$,  
Near-UV, Far-UV, {\it R} band and photometric redshift (spectroscopic 
redshift in bold if available).
Units of the X-ray fluxes are  $10^{-15}$erg s$^{-1}$cm$^{-2}$.} 
\begin{tabular}{lcccccrrcr}\hline
Name &  Separation &F$_{0.2-2\, keV}$  & F$_{2-10\, keV}$ & Log L$_{0.2-2\, keV}$  & Log L$_{2-10\, keV}$& Near-UV & Far-UV & {\it R} 
& z\ \ \ \ \ \ \ \ \  \\
\    & (\arcsec)  & \    & \ & (erg s$^{-1}$) &  (erg s$^{-1}$)   &  micro  Jy     & micro  Jy     & {\small (mag.)} & \\   
\hline
CDFS017&   0.47&      1.08 &      2.59   &      42.60&      42.98  &   0.99 &    0.30 &  21.42 &   0.724$\pm$   0.045\\
CDFS065&   1.19& $<$  0.09 &      1.14   &  $<$ 40.54&      41.64  &   1.32 &    0.90 &  19.22 &   {\bf 0.310}            \\
CDFS073&   0.80&      0.10 & $<$  0.76   &      40.98& $<$  41.86  &   1.07 &    0.66 &  19.36 &   0.440$\pm$   0.013\\
CDFS078&   2.93&      1.88 &      3.94   &      41.29&      41.61  &   8.97 &    4.33 &  19.58 &   0.180$\pm$   0.009 \\ 
CDFS088&   0.82&      2.48 &     18.73   &      42.74&      43.62  &   3.85 &    0.74 &  20.16 &   {\bf 0.605}\\
CDFS129&   0.14&      0.22 & $<$  0.94   &      40.61& $<$  41.24  &   1.61 &    1.37 &  19.18 &   {\bf 0.229}\\
CDFS132&   0.93& $<$  0.06 & $<$  0.32   &  $<$ 40.06& $<$  40.78  &   6.19 &    4.24 &  20.37 &   {\bf 0.232} \\
CDFS149&   0.50&      0.23 &      0.75   &      40.06&      40.57  &   4.67 &    2.97 &  20.12 &   {\bf 0.131}\\
CDFS152&   0.56&      0.06 & $<$  0.26   &      40.13&  $<$ 40.76  &  10.28 &    7.66 &  19.89 &   {\bf 0.248}   \\
CDFS158&   1.04&      0.09 & $<$  0.33   &      40.71&  $<$ 41.28  &   3.32 &    0.93 &  20.71 &   {\bf 0.363}\\
CDFS167&   0.84&      0.09 & $<$  0.58   &      41.24&  $<$ 42.05  &   1.73 &    0.52 &  20.52 &   {\bf 0.577}\\
CDFS185&   1.02&      0.06 & $<$  0.42   &      40.93&  $<$ 41.77  &   2.80 &    1.37 &  22.20 &   {\bf 0.511}\\
CDFS189&   0.88&      0.81 & $<$  0.34   &      40.07&  $<$ 39.70  &  62.46 &   41.54 &  16.55 &   0.075$\pm$  0.01\\
CDFS192&   0.90&      0.57 &      0.71   &      40.05&      40.14  &  35.35 &   23.69 &  17.06 &   0.086$\pm$   0.012\\
CDFS196&   0.68&      0.07 & $<$  0.37   &      41.31&  $<$ 42.03  &   1.40 &    0.25 &  21.46 &   {\bf 0.667}\\
CDFS207&   0.81&      0.18 &      0.93   &      39.83&      40.54  &   4.75 &    3.16 &  17.08 &   0.115$\pm$  0.005\\
CDFS213&   4.90(3.02)&  0.12 &    1.50   &      39.00&      40.10  &  30.49 &   22.34 &  16.19 &   0.058$\pm$   0.003  \\
CDFS228&   0.52&      0.15 & $<$  1.21   &      39.88&  $<$ 40.79  &   2.50 &    0.61 &  18.23 &   0.132$\pm$  0.01\\
CDFS236&   0.47&      0.08 & $<$  0.31   &      40.92&  $<$ 41.51  &   5.57 &    2.70 &  20.97 &   {\bf 0.456} \\
CDFS238&   0.76&      0.06 & $<$  0.29   &      40.10&  $<$ 40.78  &   1.55 &    1.02 &  20.10 &   {\bf 0.241}  \\
CDFS240&   0.55&      1.98 &     37.83   &      41.86&      43.14  &   2.91 &    0.65 &  18.86 &   0.304$\pm$  0.003\\
CDFS265&   1.35&      0.14 & $<$  0.60   &      41.19&  $<$ 41.82  &   6.04 &    1.51 &  19.47 &   0.467$\pm$  0.008\\
CDFS267&   0.39&      0.11 &      0.21   &      39.30&      39.58  &   9.54 &    7.86 &  20.27 &   0.083$\pm$   0.005\\
CDFS270&   0.96&      0.89 & $<$  0.85   &      40.52&  $<$ 40.50  &   8.25 &    2.40 &  16.07 &   0.115$\pm$  0.005\\
CDFS291&   0.94&      0.22 & $<$  0.64   &      39.83&  $<$ 40.29  &  22.84 &   16.62 &  17.11 &   {\bf 0.105} \\
CDFS292&   0.45&      0.22 & $<$  0.72   &      41.11&  $<$ 41.63  &  20.97 &   13.28 &  19.87 &   {\bf 0.366}\\
CDFS294&   0.25&      0.23 & $<$  0.81   &      39.95&  $<$ 40.50  &  17.21 &   15.24 &  18.03 &   0.117$\pm$  0.006\\
CDFS304&   0.92&      1.00 &      9.23   &      41.93&      42.90  &   1.41 &    0.88 &  20.35 &   {\bf 0.424}  \\
CDFS325&   3.57(1.4)&    1.45 &   5.62   &      41.87&      42.46  &   5.27 &    2.81 &  19.88 &   {\bf 0.347}   \\
\hline
\end{tabular}
\footnotesize $^1$ CDFS sources have been detected in at least one of the seven bands defined in Alexander et al 2003 (see text).
\end{center}
\end{table*}

\begin{table*}
\begin{center}\caption{\label{Tab:CDFS-Optical} 
Optical magnitudes of the selected objects. }
\begin{tabular}{lccccc}\hline
Name   & I & R & V  & B  & U  \\
\hline
CDFS017 &  20.205 &  21.294 &  22.335 &  22.785  & 22.390 \\
CDFS065 &  18.147 &  19.112 &  20.104 &  21.197  & 21.566 \\
CDFS073 &  18.255 &  19.258 &  20.525 &  21.511  & 22.229 \\
CDFS078 &  18.851 &  19.440 &  19.923 &  20.522  & 20.457 \\
CDFS088 &  19.103 &  20.069 &  21.112 &  21.622  & 21.204 \\
CDFS129 &  18.308 &  19.064 &  19.822 &  20.799  & 21.122 \\
CDFS132 &  19.809 &  20.273 &  20.722 &  21.312  & 21.029 \\
CDFS149 &  19.433 &  20.004 &  20.529 &  21.109  & 21.081 \\
CDFS152 &  19.376 &  19.784 &  20.221 &  20.774  & 20.355 \\
CDFS158 &  19.975 &  20.583 &  21.252 &  22.106  & 21.688 \\
CDFS167 &  19.373 &  20.401 &  21.663 &  22.392  & 22.144 \\
CDFS185 &  22.012 &  22.130 &  22.811 &  23.069  & 22.199 \\
CDFS189 &  15.875 &  16.420 &  16.986 &  17.464  & 17.539 \\
CDFS192 &  16.328 &  16.905 &  17.499 &  18.013  & 18.136 \\
CDFS196 &  20.823 &  21.363 &  22.175 &  22.516  & 22.043 \\
CDFS207 &  16.230 &  16.949 &  17.642 &  18.490  & 18.972 \\
CDFS213 &  15.240 &  16.021 &  16.770 &  17.395  & 17.803 \\
CDFS228 &  17.371 &  18.073 &  18.758 &  19.548  & 19.883 \\
CDFS236 &  20.325 &  20.843 &  21.471 &  21.913  & 21.120 \\
CDFS238 &  19.198 &  20.023 &  20.809 &  21.680  & 21.805 \\
CDFS240 &  17.919 &  18.710 &  19.466 &  20.550  & 20.878 \\
CDFS265 &  18.647 &  19.330 &  20.246 &  20.845  & 20.457 \\
CDFS267 &  19.826 &  20.136 &  20.320 &  20.798  & 20.535 \\
CDFS270 &  15.190 &  15.934 &  16.637 &  17.505  & 18.027 \\
CDFS291 &  16.332 &  16.971 &  17.605 &  18.283  & 18.549 \\
CDFS292 &  19.467 &  19.756 &  20.131 &  20.597  & 19.738 \\
CDFS294 &  17.264 &  17.895 &  18.482 &  19.143  & 19.301 \\
CDFS304 &  19.309 &  20.223 &  21.321 &  22.175  & 22.176 \\
CDFS325 &  18.937 &  19.777 &  20.653 &  21.267  & 21.193 \\
\hline
\end{tabular}
\end{center}
\end{table*}

\section{Extinction-Corrected UV Fluxes from Template Fitting}
\label{Uvfluxes}

To obtain the intrinsic UV flux at rest frame 
2000~\AA\, from the optical spectra and UV observed values 
we adopted a template fitting approach (Bolzonella, Miralles, \& 
Pell\'o 2000; Babbedge et al. 2004).
We used 5 different galaxy templates calculated with the GRASIL code
(Silva et al. 1998) that are accessible from a  
library of galaxy models
\footnote{http://web.oapd.inaf.it/granato/grasil/modlib/modlib.html}. 
The extinction--free templates represent spirals (Sa, Sb and Sc), and 
the starburst galaxies M82 and Arp220. 
    Notice that, for  spirals, the obtained galaxy type
    represents an average of the observed spectral energy distribution,
    but due to the large dispersion present in most of the observables
    (e.g.~absolute magnitudes, colours) and in derived
    physical properties (like star formation history)
    we can not unequivocally assign a morphological type 
    to a single spectrum.
    Therefore, the galaxy templates just correspond to an average of
    the intrinsic stellar emitted light without extinction; we do not
    attempt to fit as well a galaxy type.

From the comparison of models and observations we obtained the
best match using a $\chi^2$ minimization. The observed data included the five 
optical bands (Table~\ref{Tab:CDFS-Optical}) provided 
by Wolf et al. (2004) plus the two UV GALEX bands.
For each object the original GRASIL template was shifted according to 
the  z$_{sp}$ or the  COMBO-17 photometric redshift. The observed flux 
F$_{o}(\lambda$) and the flux given by the template, 
F$_{t}(\lambda$) can be expressed as a function of the extinction as  
(e.g. Osterbrook 1989)

\begin{equation}
F_{o}(\lambda) = F_{t}(\lambda) \times 10^{-0.4 {\rm A_V}\, k(\lambda)/R}
\end{equation}
where the values $k(\lambda)$ and $R$ are tabulated for different
extinction curves. It is known that the appropriate extinction curve for 
local galaxies depends on the type of galaxy and on parameters, such as 
metallicity, that presumably determine the characteristics of the dust. 
These parameters are not known for the selected sample, so we included the 
extinction law as one of the unknowns in the fitting grid.
The best fit was found in a three dimensional grid 
where we change the galaxy template, the value of the visual extinction (A$_V$)
and the extinction curve. 
For this analysis we adopted three different extinction curves:
Milky Way (MW; Seaton 1979, Howarth 1983), Large Magallanic Cloud (LMC; Howarth 1983) and 
Calzetti law for starburst galaxies (Calzetti; Calzetti, Kinney and 
Storchi-Bergmann 1994).
Unfortunaly there is no independent information to further
constrain  the extinction (e.g. hydrogen emission line
ratios) or the morphologies of the galaxies (the optical images 
provided 
by COMBO-17 are not good enough for a detailed morphological  study).

\begin{figure*}
\setlength{\unitlength}{1cm}           
\begin{picture}(14,22)       
\put(-1.5,11.){\includegraphics{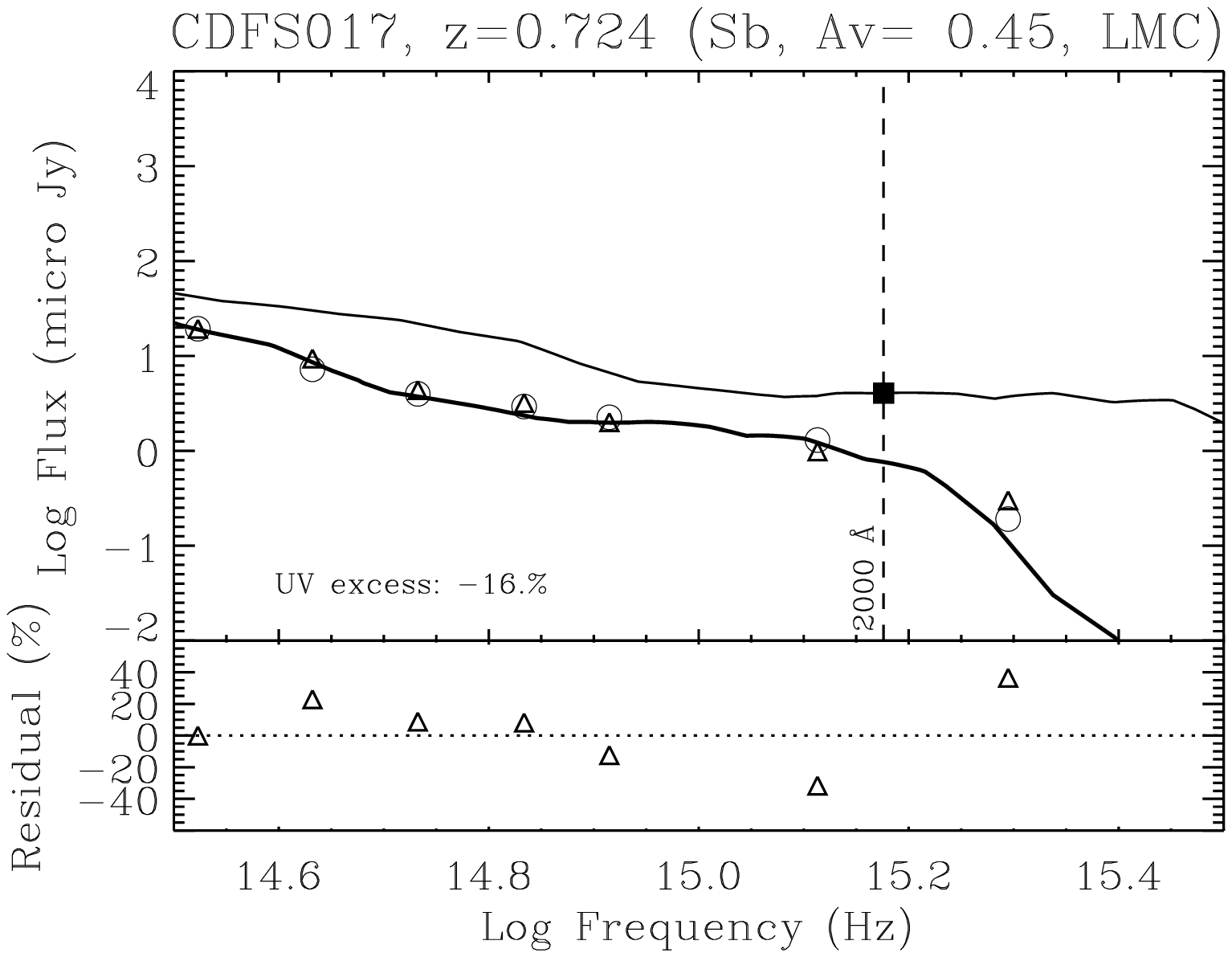}}
\put( 6.5,11.){\includegraphics{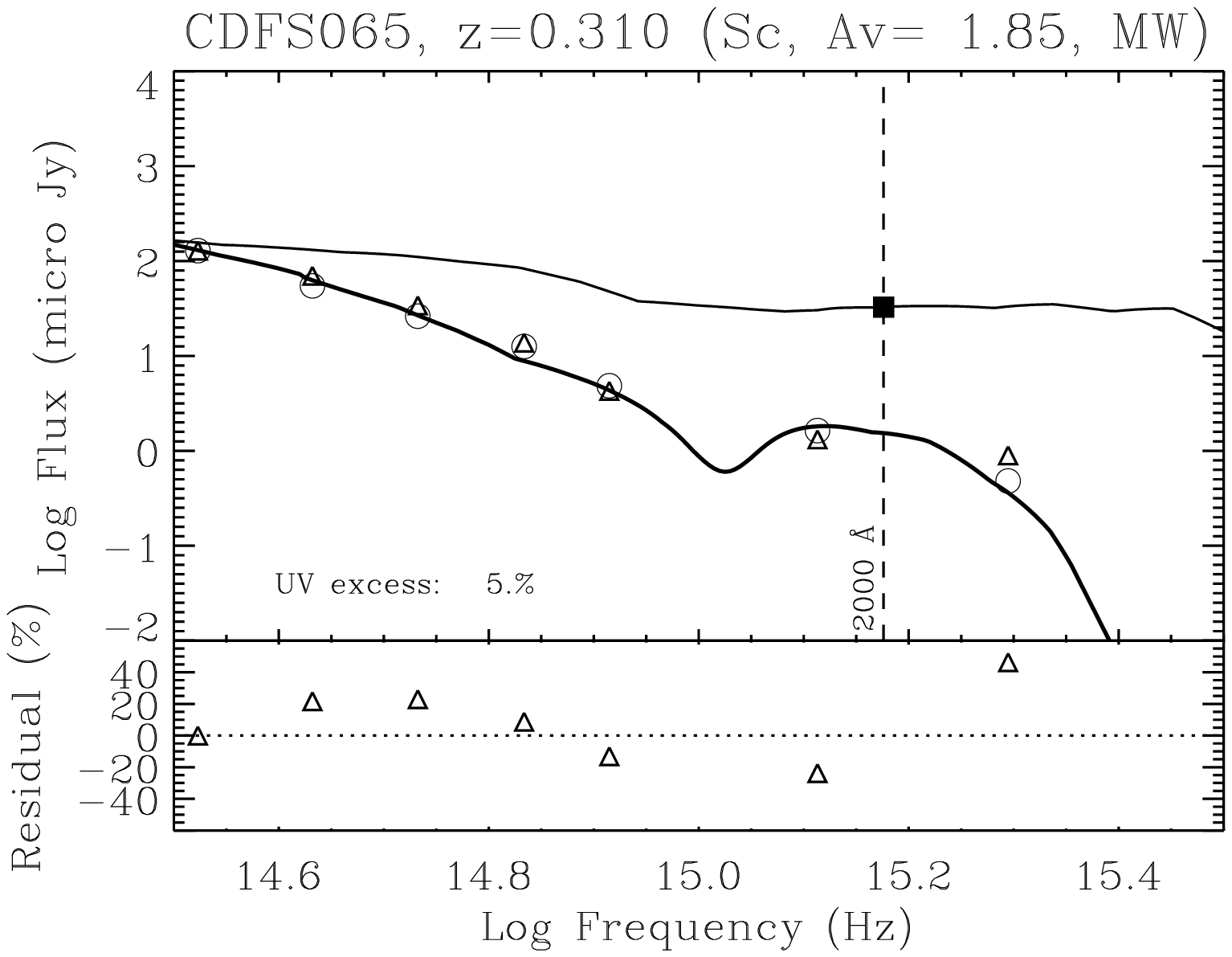}}
\put(-1.5, 4.5){\includegraphics{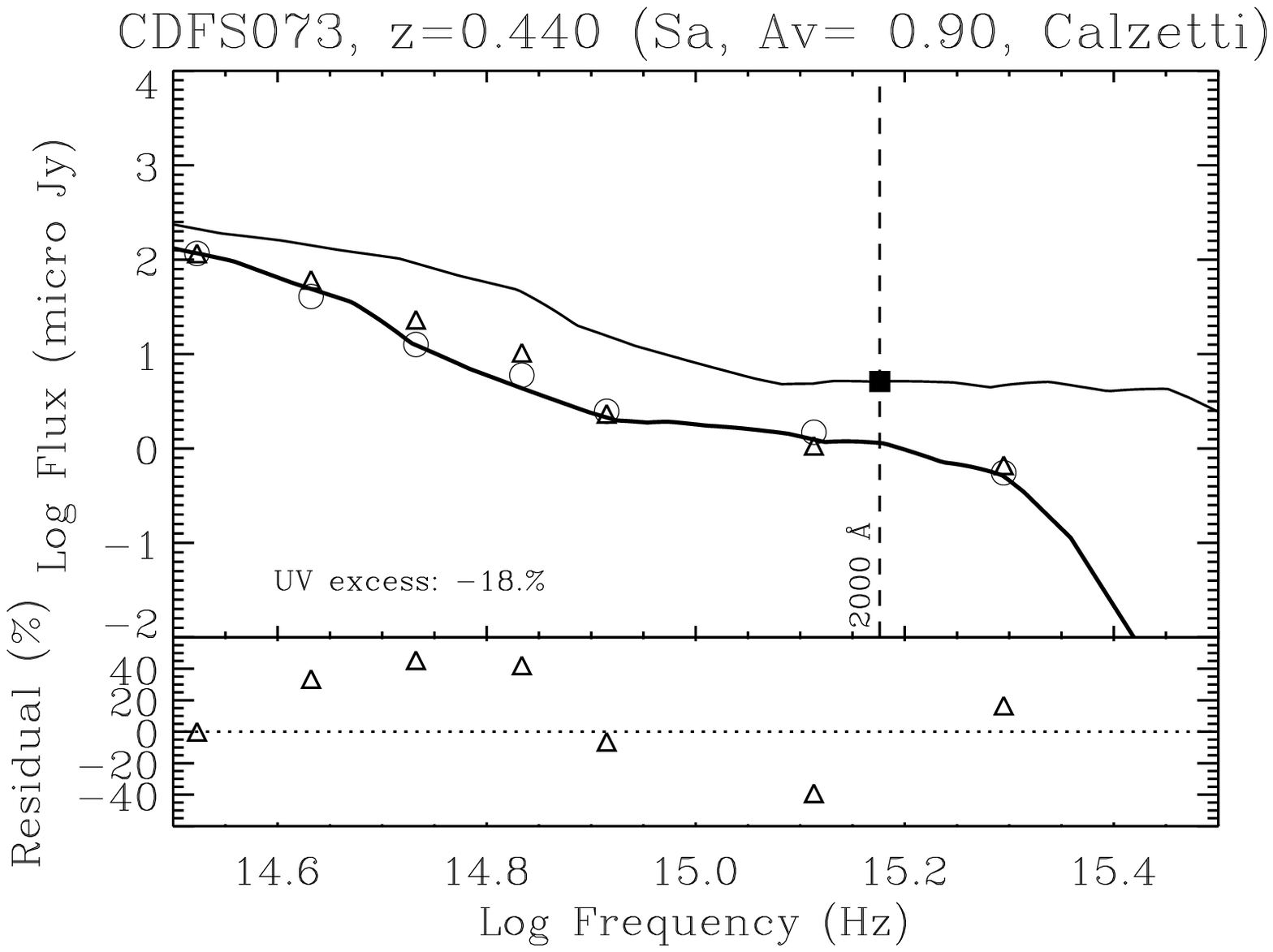}}
\put( 6.5, 4.5){\includegraphics{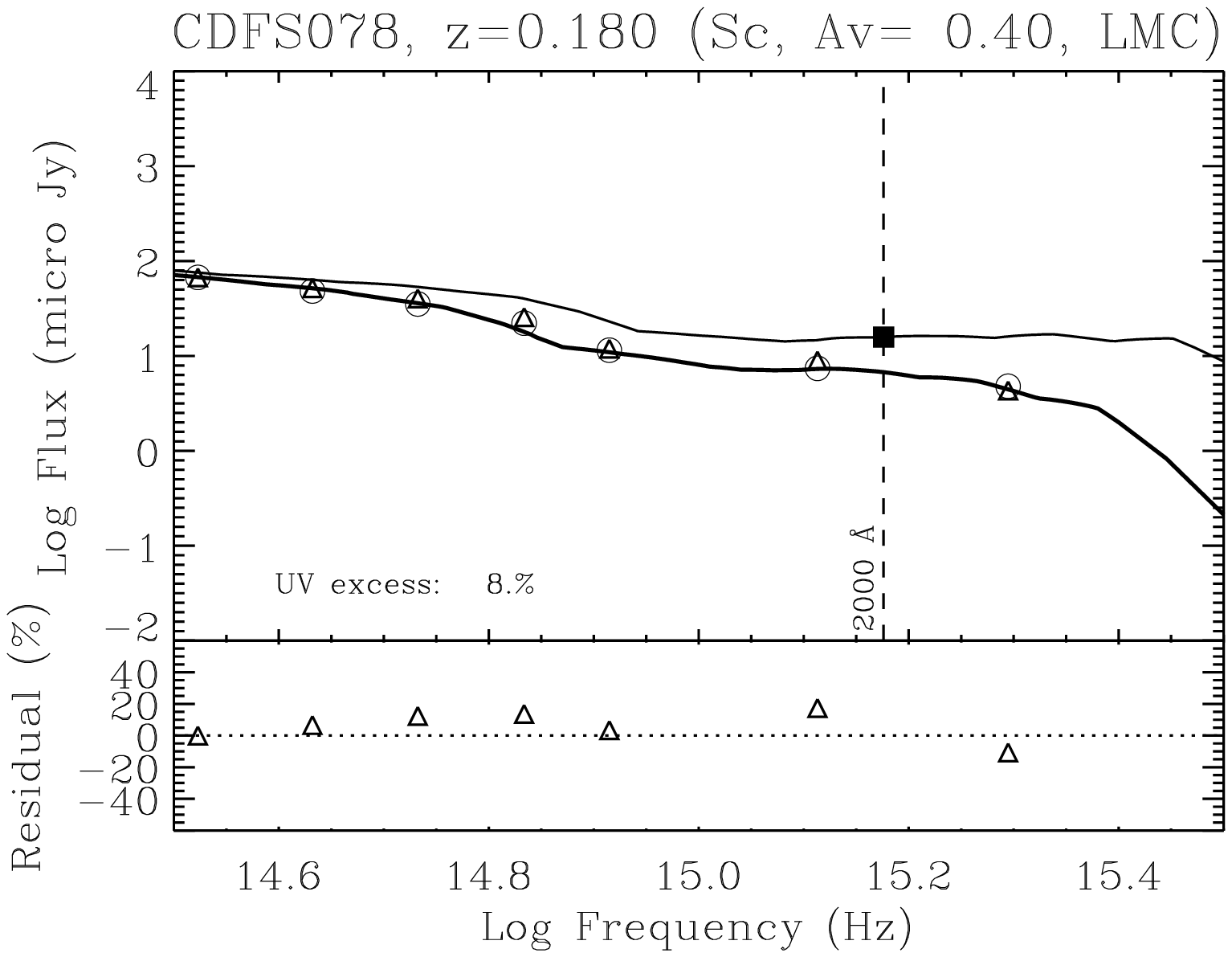}}
\put(-1.5,-2.0){\includegraphics{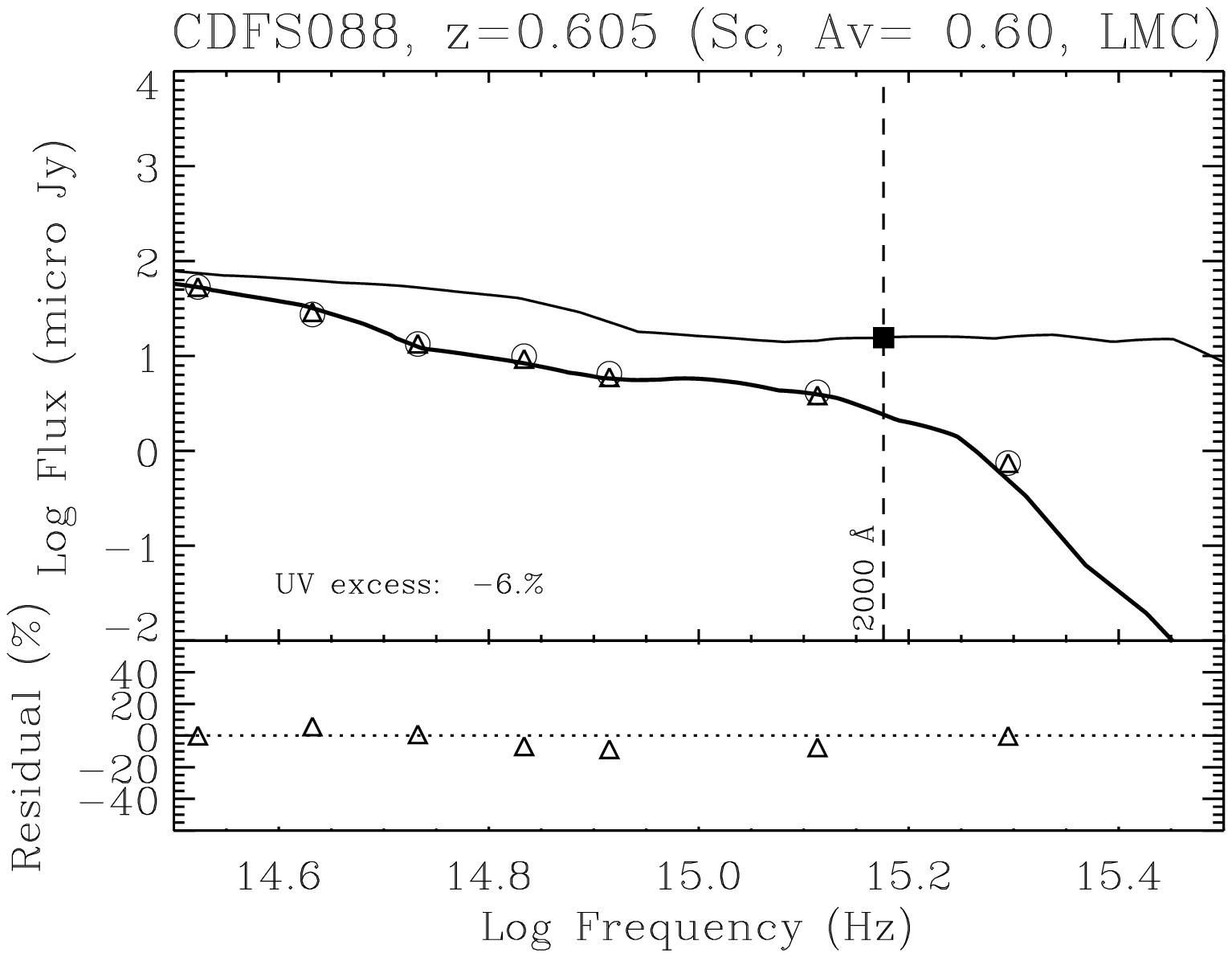}}
\put( 6.5,-2.0){\includegraphics{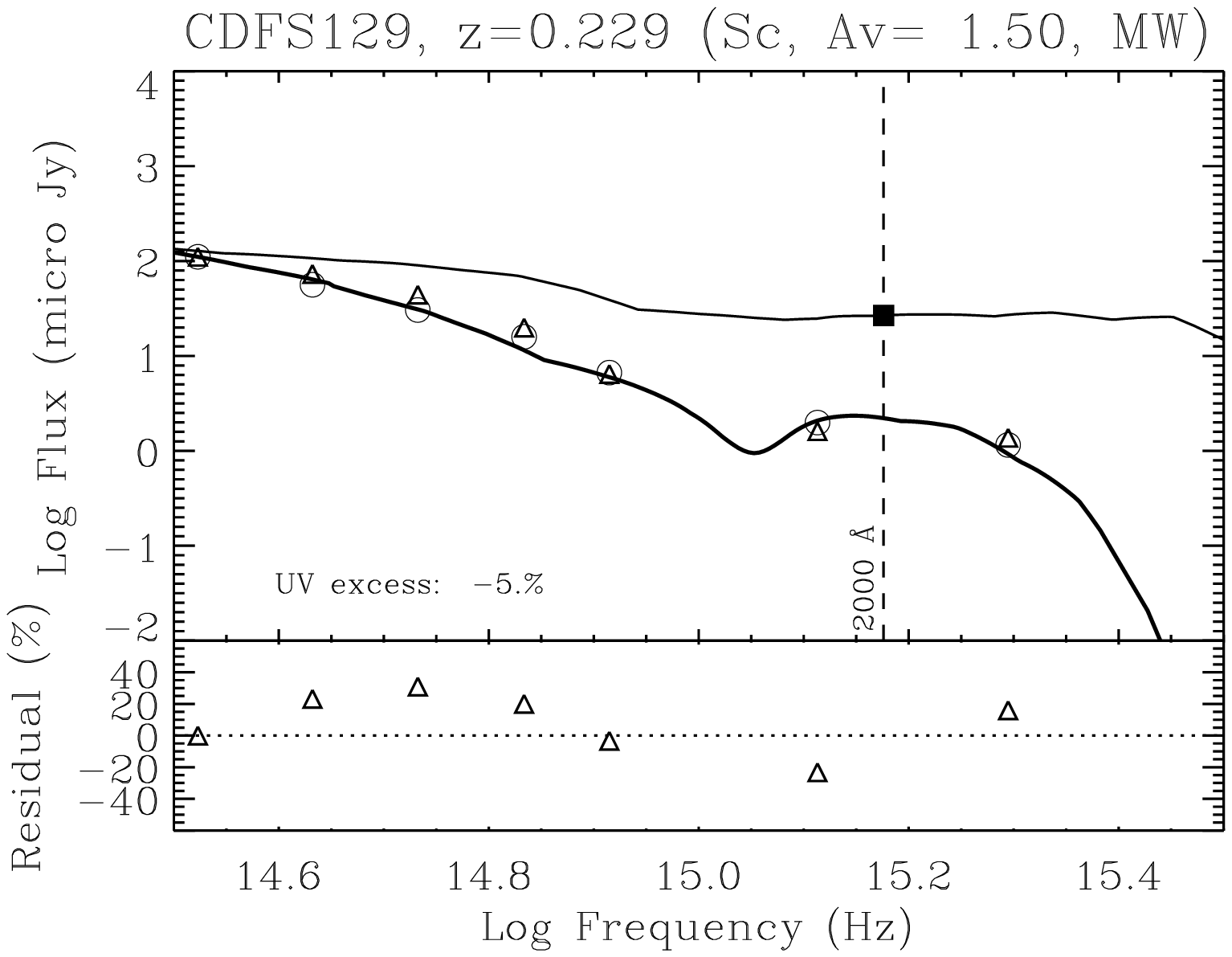}}
\end{picture}
\caption{\label{SEDs_1} 
SEDs of the selected galaxies. Triangles are the observed values; 
circles are the fluxes extracted from the modelled spectra  
redshifted and without extinction correction 
(thick line). The thin line is the intrinsic template at rest frame 
without extinction. 
The black square is the rest frame flux at 2000\AA. On top of
each panel is the redshift, and the result of the best fit
(i.e. galaxy template, visual extinction and extinction law).  
     The UV excess  (bottom left corner of the panel)
     represents the difference  between
     the observed near and far UV fluxes  and the corresponding
     values given by the best fit model and is defined in section~\ref{Uvfluxes}).
}
\end{figure*}

The results of the fitting are presented in Table~\ref{Tab:CDFS-Templates} and 
the individual fitted spectra compared to observed values 
are plotted in Figure~\ref{SEDs_1}.
Once the best fit is obtained we calculate 
the extinction corrected UV flux at 2000~\AA\ in the corresponding
rest frame template (Table~\ref{Tab:CDFS-Templates}).
The provided UV flux corresponds to the rest frame value at 2000~\AA\, 
and it is directly obtained from the dust free template taking into account 
the redshift of the source.
A UV excess, that quantifies the deviation of the model from the observed 
UV values, is calculated by adding the difference  between 
the observed near and far UV fluxes  and the corresponding 
values given by the best fit model. The obtained quantities, normalized by 
the   
sum of the observed UV fluxes and multiplied by 100,
are around $\pm$15\% and are indicated as labels to the individual panels
in Figure~\ref{SEDs_1}.


The rescaling of the original template 
to the observed flux in the I band, gives an estimate of the stellar content 
of the galaxies  (Table~\ref{Tab:CDFS-Templates}). However,  as
the I band could be contaminated by the emission of massive 
stars (red supergiants) which do not contribute significantly to the total 
mass of the galaxy, the given 
stellar mass should be treated as an upper limit of the total stellar mass.


\begin{table*}
\begin{center}\caption{\label{Tab:CDFS-Templates} 
Best template fit, estimated stellar mass (M$_\star$) and 
derived UV flux  at 2000\AA\ (rest frame and extinction corrected). 
F2000* is the corrected flux combining optical,  UV and Spitzer fluxes. 
Column 7 gives the derived SFR(UV)  
(in parenthesis,  using the  F2000* values). Column 8 is the SFR(X)
(in parenthesis the SFR given by the hard X-rays for galaxies  
probably harbouring an obscured AGN).
Column 9 represents the percentage of the SFR(X) that could be contaminated by 
the luminosity of LMXB.
}
\begin{tabular}{lccccccrrcc}\hline
Name   & Template &  M$_\star$   &Extinction  &     A$_V$         & log F2000  & log F2000$^*$    & SFR(UV) & SFR(X)& LMXB & Notes \\
       &          & ($\times 10^{10}$\Msolar)  & Curve    &   (mag.)       &  (micro Jy)& (micro Jy)  & (\Myear)& (\Myear)&\%   &   \\
\hline
CDFS017&  Sb           &  14    &      LMC  & 0.45 &  0.61 &   --  &  13          &   866{\bf(1888)}       &  0.18   & (a)  \\
CDFS065&  Sc           &   5.6  &       MW  & 1.85 &  1.51 & 1.42  &  14          &     7.6                &  8.0    &   \\
CDFS073&  Sa           &  39    & Calzetti  & 0.90 &  0.71 & 0.90  &   5.1        &    21                  & 20      &   \\
CDFS078&  Sc           &   0.87 &      LMC  & 0.40 &  1.20 & 1.08  &   2.0        &    43{\bf(82)}         &  0.22   & (a)   \\
CDFS088&  Sc           &  11    &      LMC  & 0.60 &  1.19 & 1.60  &  33          &  1212{\bf(8321)}       &  0.10   & (a)   \\
CDFS129&  Sc           &   2.4  &       MW  & 1.50 &  1.43 & 1.83  &   5.9        &     8.9                &  3.0    &   \\
CDFS132&  Sc           &   0.62 &      LMC  & 0.10 &  0.83 & 0.84  &   1.5        &     2.5                &  2.7    &   \\
CDFS149&  Sc           &   0.26 &      LMC  & 0.40 &  0.95 & 1.02  &   0.57       &     2.5                &  1.1    &   \\
CDFS152&     M82-like  &   0.92 &      LMC  & 0.25 &  1.27 & 1.12  &4.9{\bf(3.4)} &     2.9                &  3.4    & (b)  \\
CDFS158&     Sc        &   1.5  &      LMC  & 0.45 &  0.80 & 1.10  &   4.0        &    11                  &  1.4    &   \\
CDFS167&  Sb           &  16    & Calzetti  & 0.95 &  0.86 & 0.68  &  14          &    39                  &  4.7    &   \\
CDFS185&  Arp220-like  &   0.13 &       MW  & 0.30 &  0.64 & 0.77  &   6.2        &    19                  &  0.08   &   \\
CDFS189&    Sc         &   2.1  &      LMC  & 0.65 &  2.36 & 2.47  &   4.4        &     2.6                &  9.0    &   \\
CDFS192&    Sc         &   1.9  &      LMC  & 0.70 &  2.18 & 2.27  &   3.9        &     2.5                &  8.3    &   \\
CDFS196&     M82-like  &   2.4  &      LMC  & 0.55 &  0.80 & 1.05  &  17.0        &    45                  &  0.58   &   \\
CDFS207&  Arp220-like  &   1.2  &       MW  & 2.85 &  2.92 & 0.94  &40{\bf(0.42)} &     1.5                &  8.5    & (b) \\
CDFS213&    Sc         &   2.2  &       MW  & 1.50 &  2.62 & 2.26  &4.7{\bf(2.1)} &     0.22               &110      & (c,d) \\
CDFS228&  Arp220-like  &   0.54 &      LMC  & 2.35 &  2.47 &   --  &  19          &     1.7                &  3.6    &   \\
CDFS236&     M82-like  &   1.5  &      LMC  & 0.20 &  0.94 & 1.54  &   9.4        &    18                  &  0.91   &   \\
CDFS238&  Sc           &   1.2  &       MW  & 1.20 &  1.07 & 1.21  &   2.9        &     2.7                &  4.7    &   \\
CDFS240&  Sc           &   6.5  &      LMC  & 1.25 &  1.60 & 2.07  &  17          &   160{\bf(2783)}       &  0.45   & (a)   \\
CDFS265&     M82-like  &   7.4  & Calzetti  & 1.30 &  1.61 & 1.62  &  46          &    34                  &  2.4    &   \\
CDFS267&  Arp220-like  &   0.02 &      LMC  & 0.55 &  1.45 & 1.40  &   0.67       &     0.44               &  0.51   &   \\
CDFS270&  Sc           &   9.7  &      LMC  & 1.95 &  2.65 & 1.57  &21{\bf(1.8)}  &    7.3                & 15       & (b)\\
CDFS291&  Sc           &   2.8  &      LMC  & 0.90 &  2.19 & 1.63  &6.1{\bf(1.7)}&    1.5                & 21        & (b)\\
CDFS292&  Arp220-like  &   0.73 & Calzetti  & 0.55 &  1.68 & 1.84  &  31          &   28                  &  0.28    &   \\
CDFS294&  Sc           &   1.5  &       MW  & 0.75 &  1.82 & --    &   3.3        &    2.0                &  8.4     &   \\
CDFS304&  Sb           &   8.0  & Calzetti  & 0.95 &  0.83 & --    &   6.1        &  189{\bf(1587)}       &  0.47    & (a)  \\
CDFS325&  Sc           &   3.5  &       MW  & 0.75 &  1.21 & --    &   9.2        &  164{\bf(577)}        &  0.23    & (a)  \\
\hline
\end{tabular}\\
Notes on individual galaxies: (a) Possible obscured AGN,
(b) galaxies in which we used the Spitzer data 
to correct the observed UV fluxes,
(c) extremely weak in X-ray and (d) high contamination due to LMXB.  
\end{center}
\end{table*}

\newpage

\setcounter{figure}{0}
\begin{figure*}
\setlength{\unitlength}{1cm}           
\begin{picture}(14,22)       
\put(-1.5,11.){\includegraphics{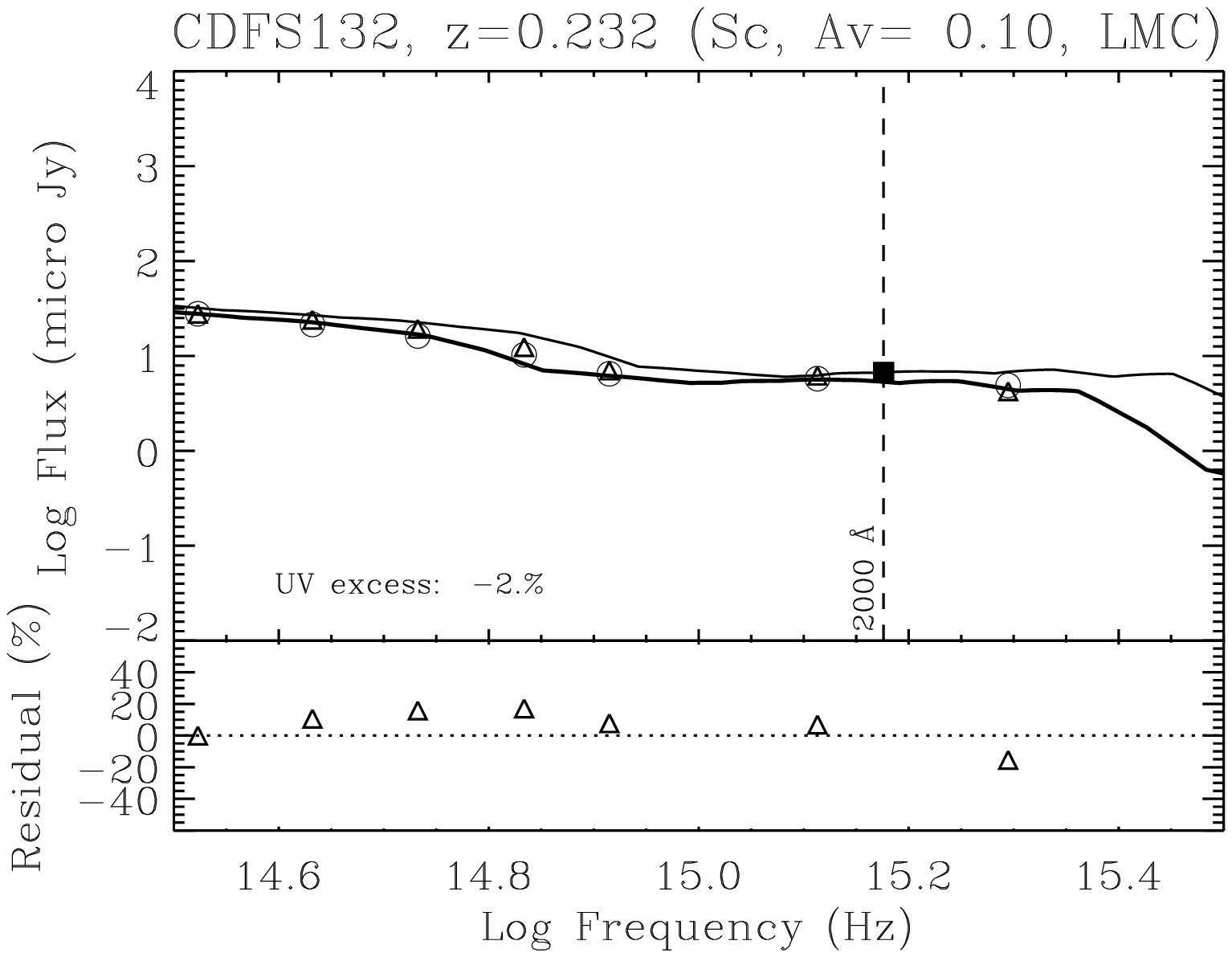}}
\put( 6.5,11.){\includegraphics{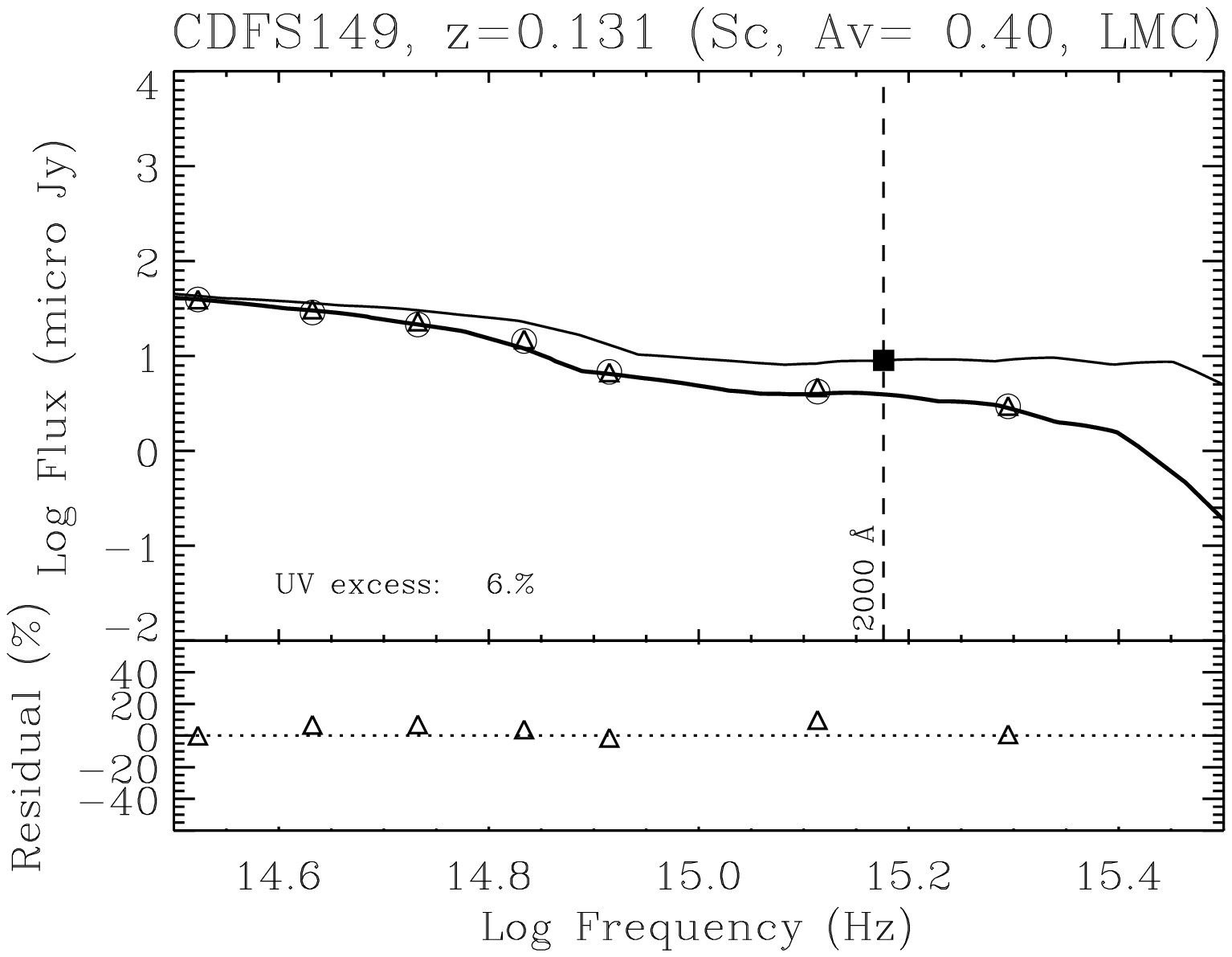}}
\put(-1.5, 4.5){\includegraphics{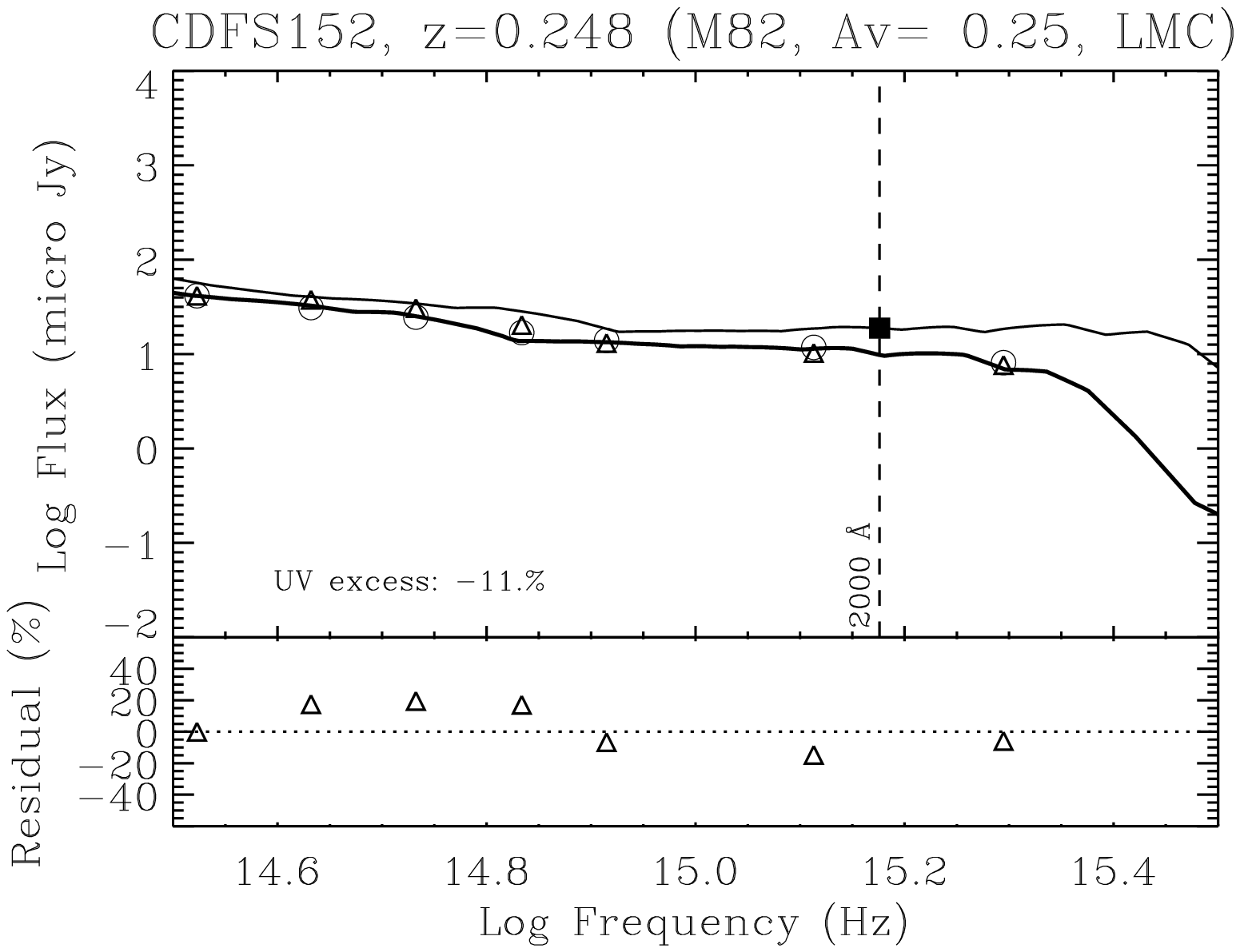}}
\put( 6.5, 4.5){\includegraphics{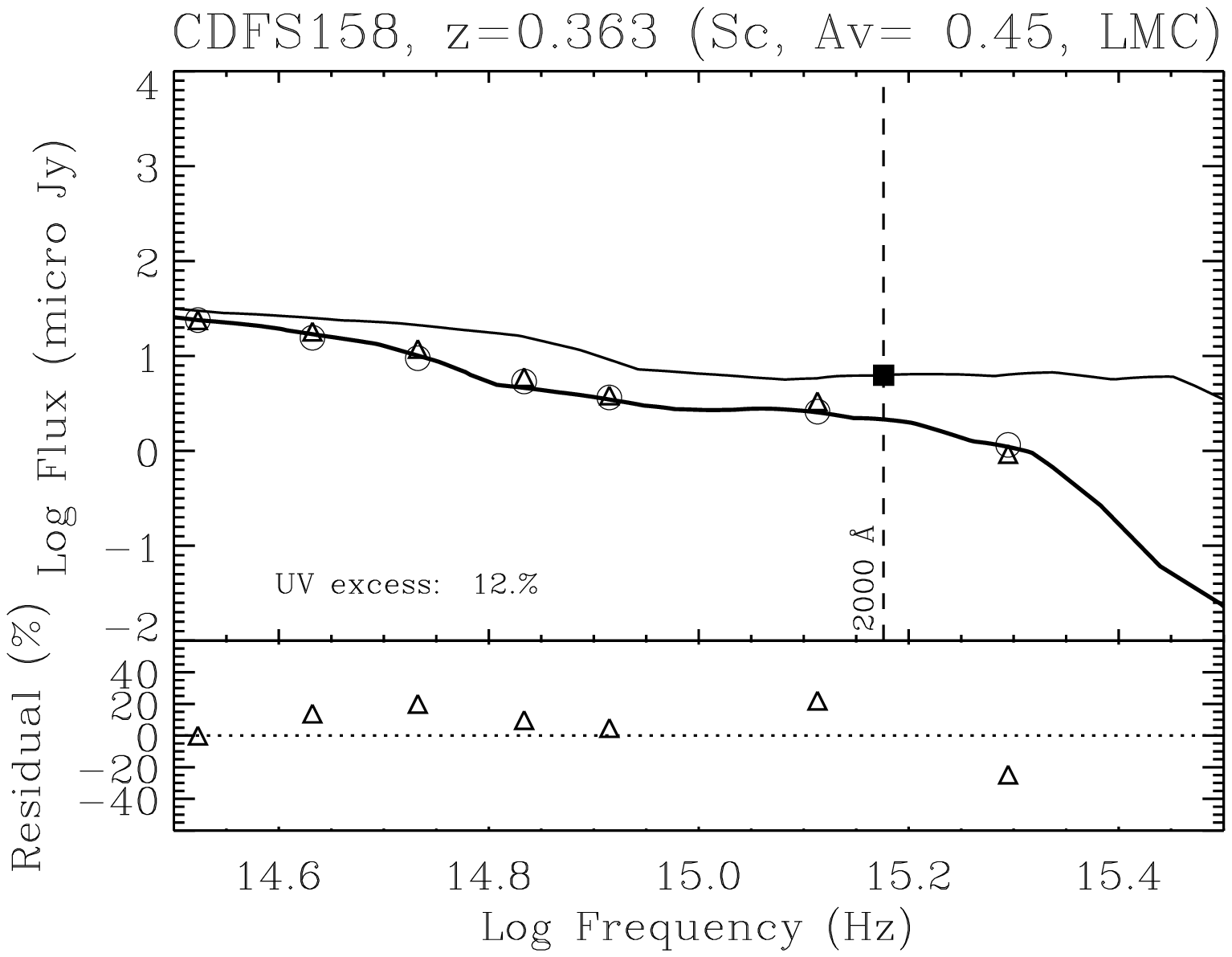}}
\put(-1.5,-2.0){\includegraphics{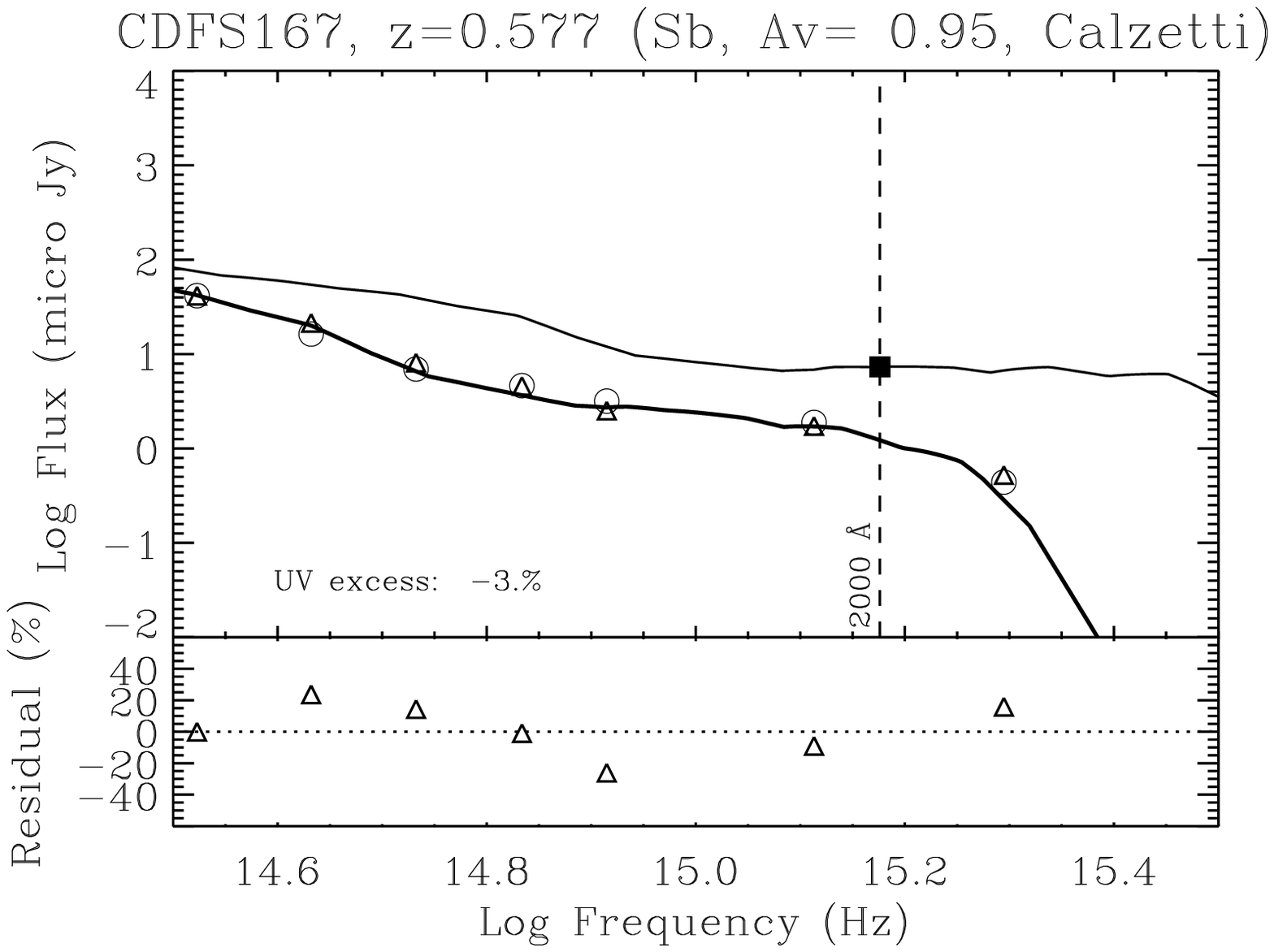}}
\put( 6.5,-2.0){\includegraphics{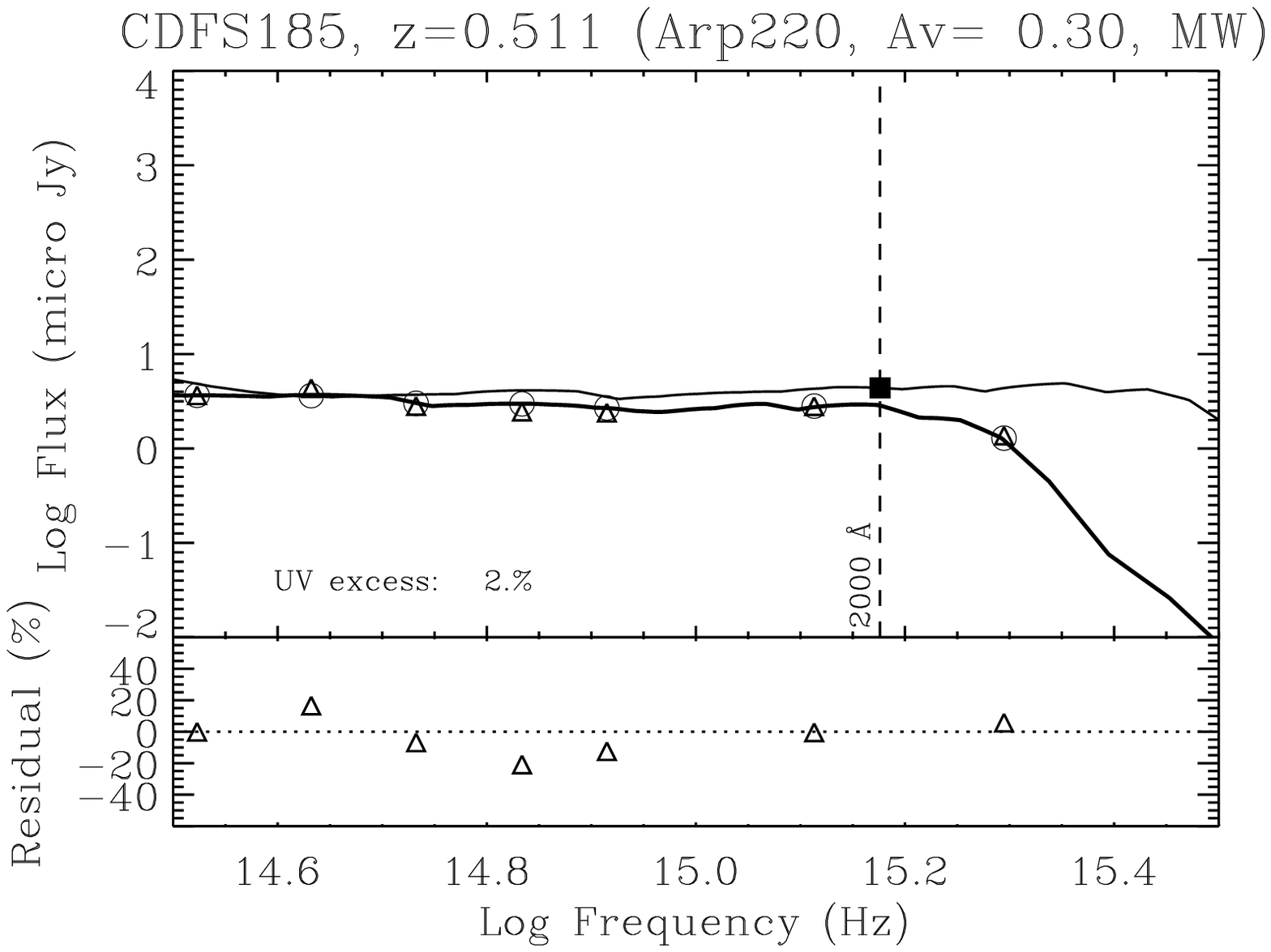}}
\end{picture}
\caption{\label{SEDs_2} 
Continued.
}
\end{figure*}

\setcounter{figure}{0}
\begin{figure*}
\setlength{\unitlength}{1cm}           
\begin{picture}(14,22)       
\put(-1.5,11.){\includegraphics{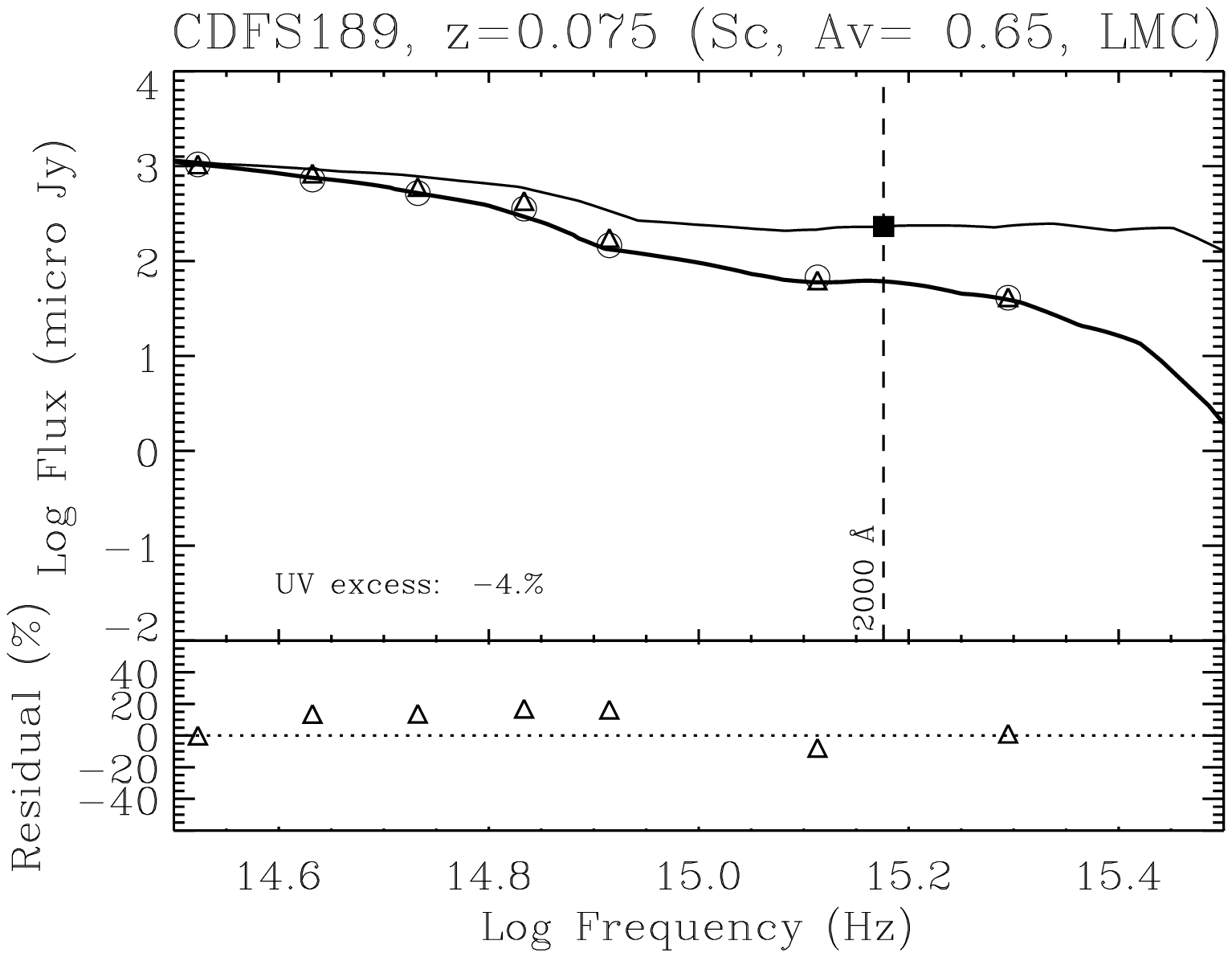}}
\put( 6.5,11.){\includegraphics{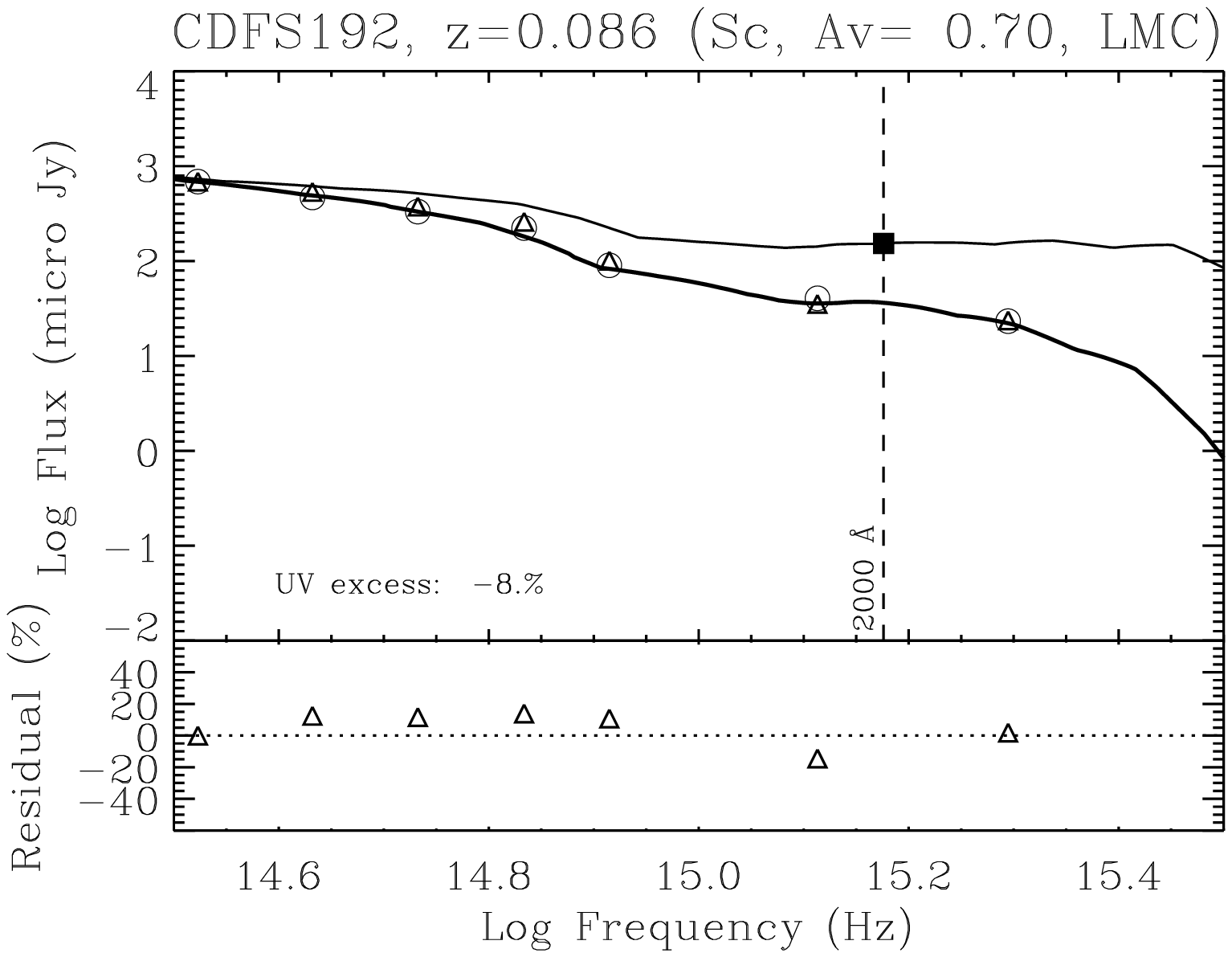}}
\put(-1.5, 4.5){\includegraphics{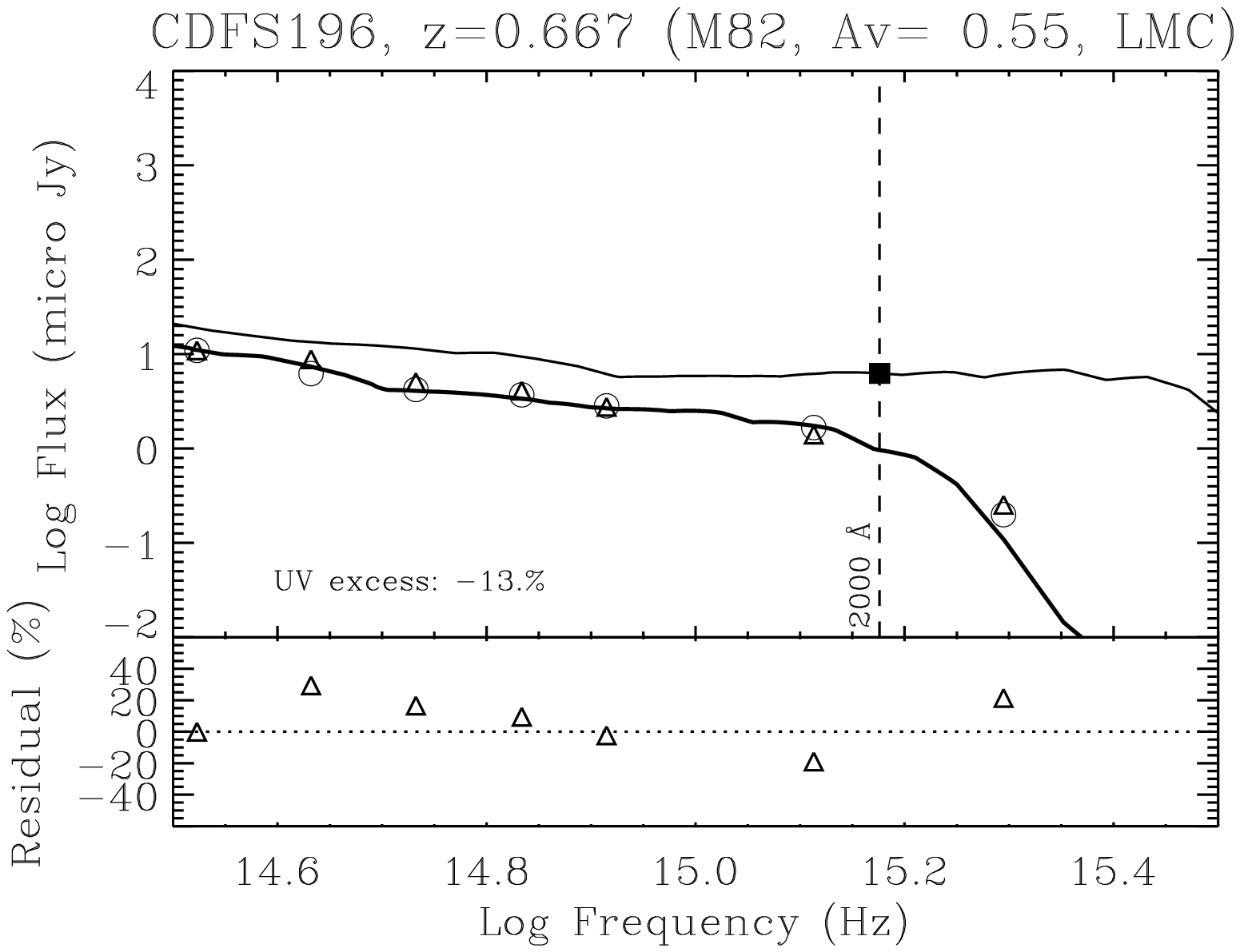}}
\put( 6.5, 4.5){\includegraphics{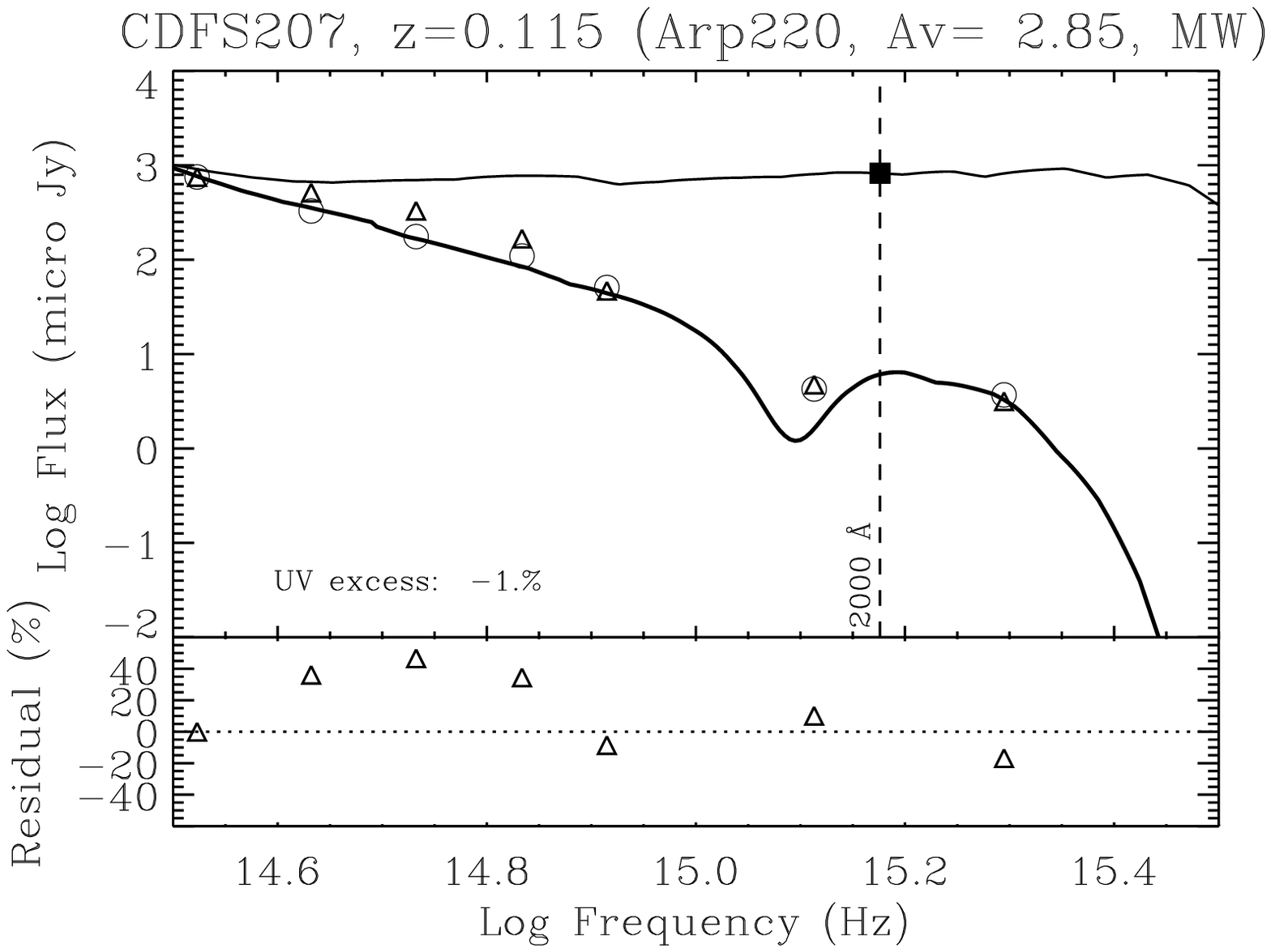}}
\put(-1.5,-2.0){\includegraphics{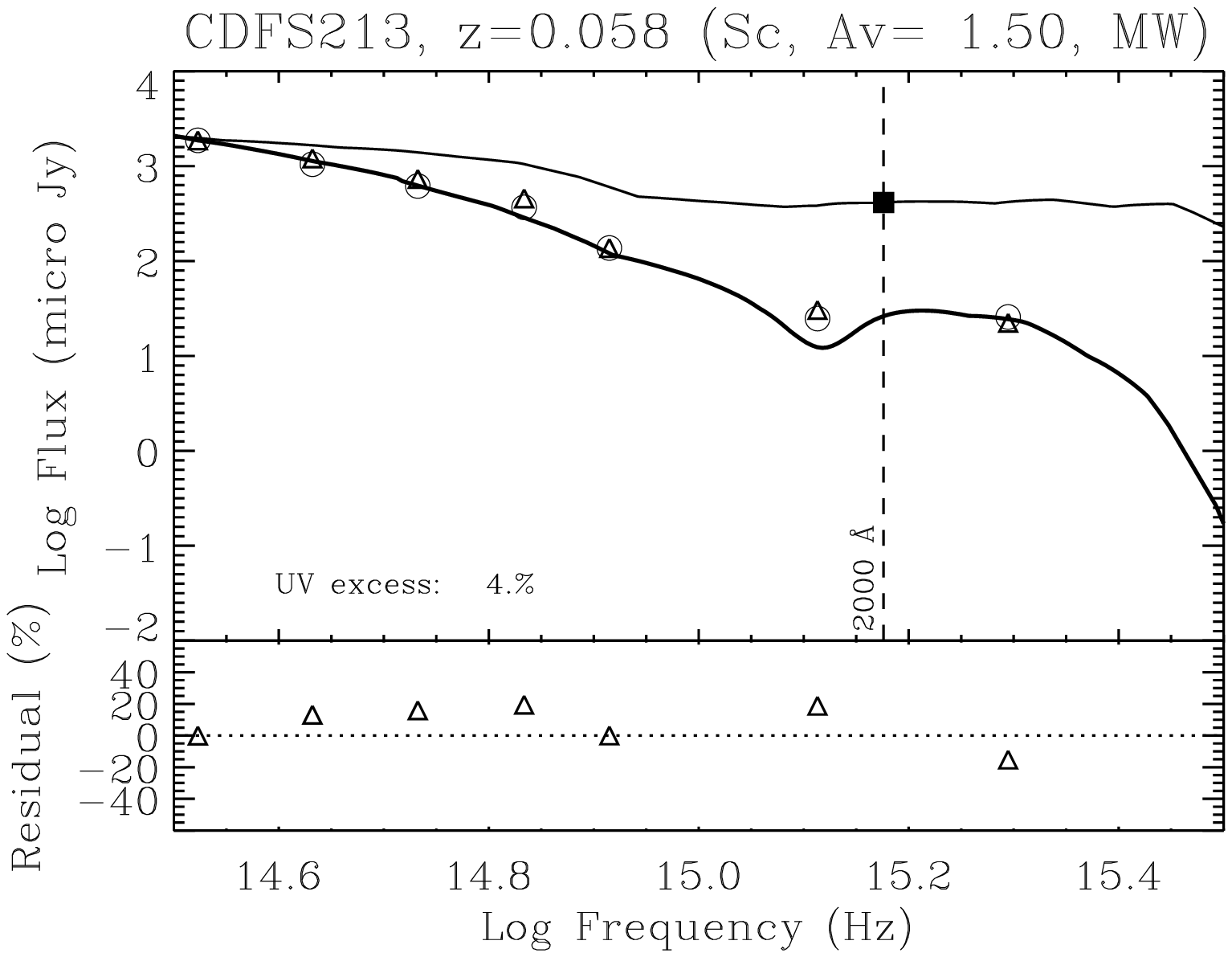}}
\put( 6.5,-2.0){\includegraphics{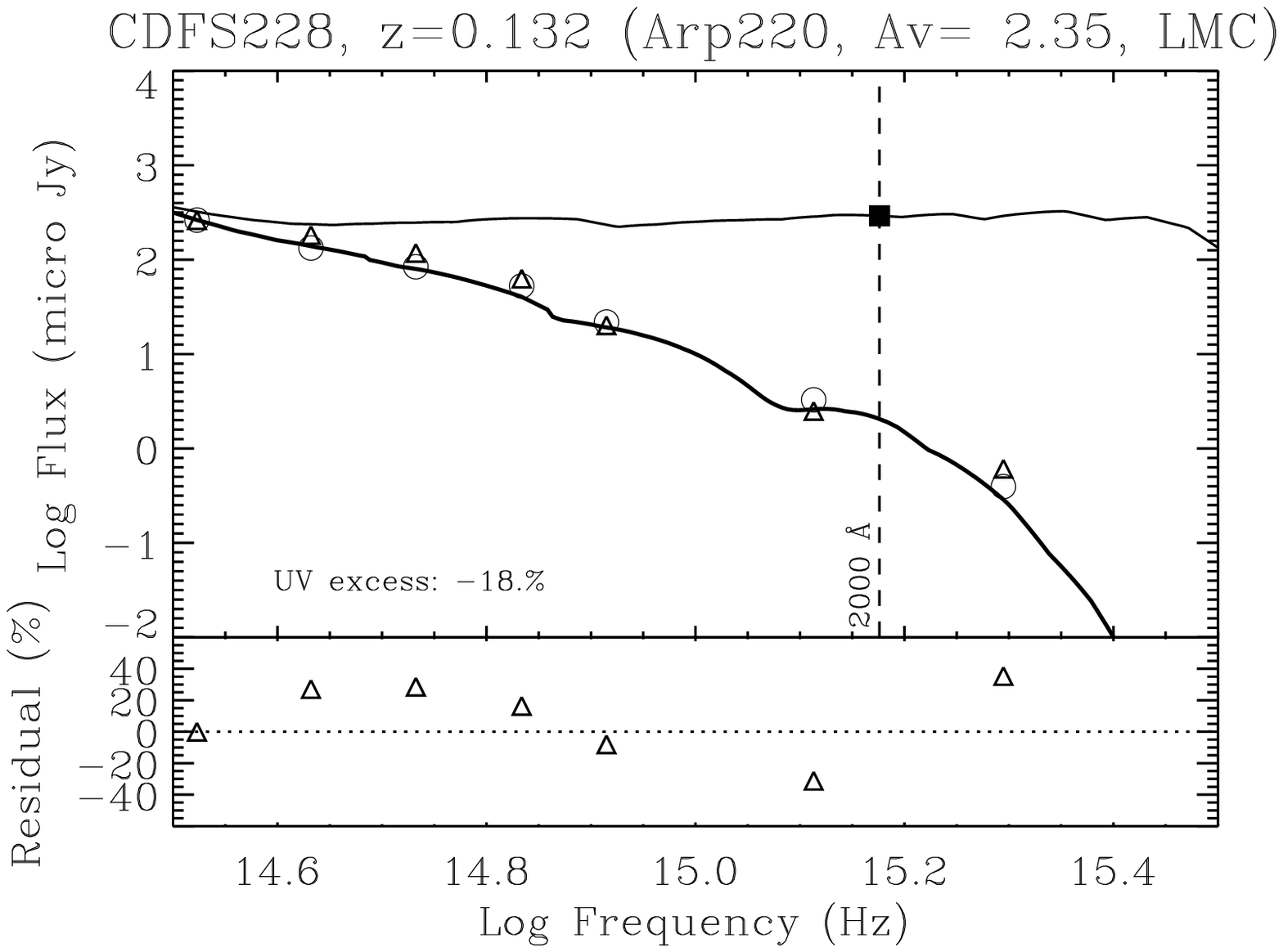}}
\end{picture}
\caption{\label{SEDs_3} Continued.
}
\end{figure*}

\setcounter{figure}{0}
\begin{figure*}
\setlength{\unitlength}{1cm}           
\begin{picture}(14,22)       
\put(-1.5,11.){\includegraphics{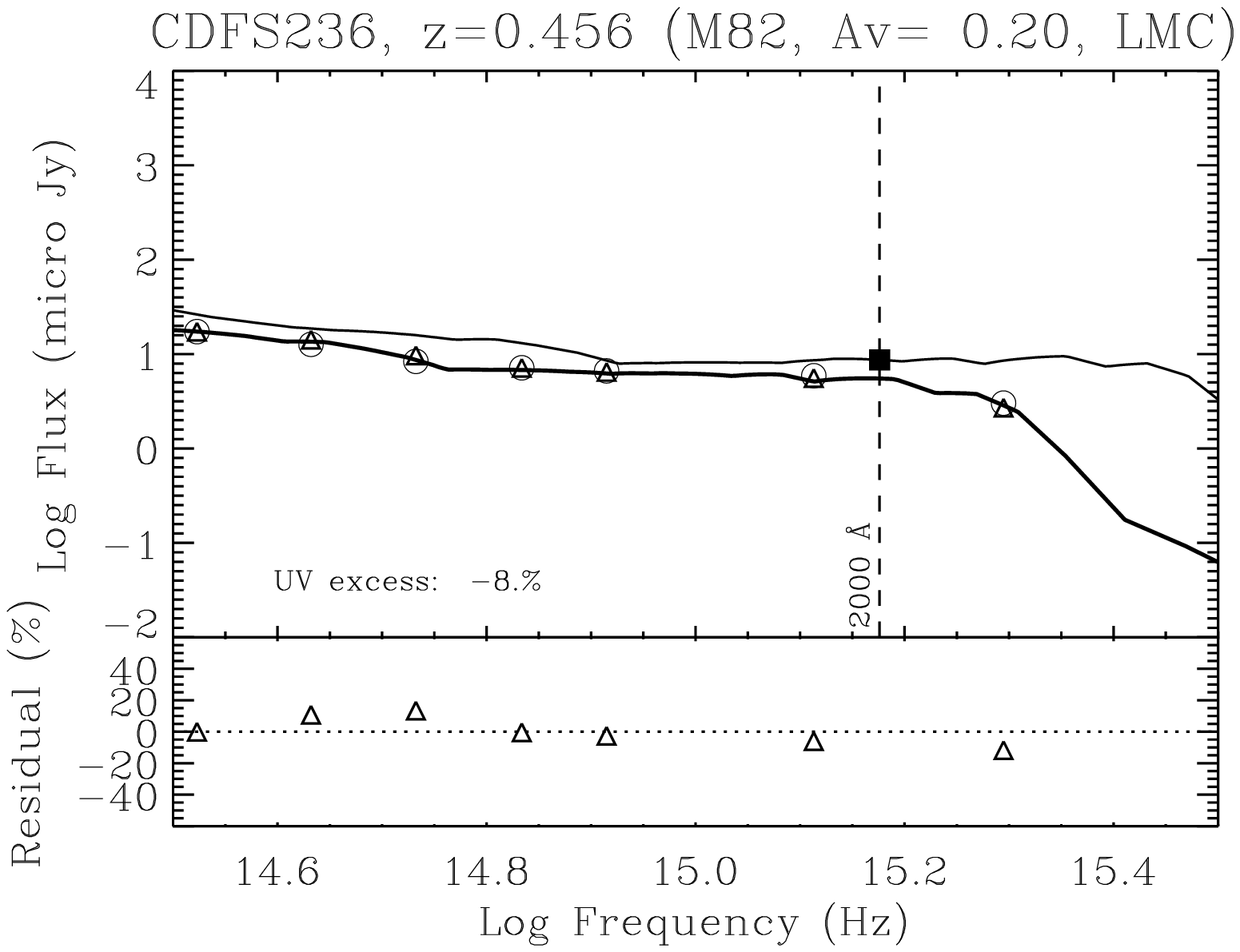}}
\put( 6.5,11.){\includegraphics{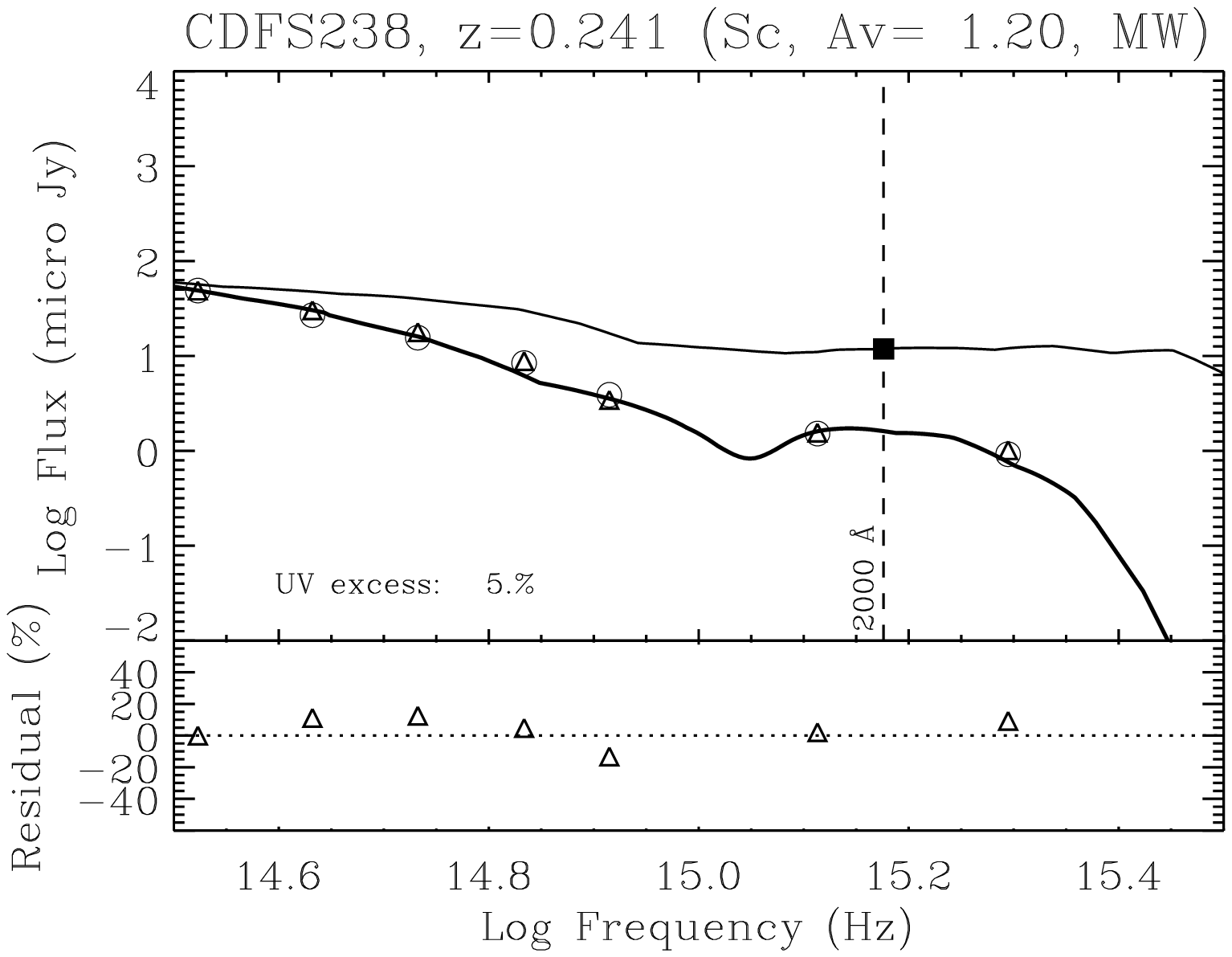}}
\put(-1.5, 4.5){\includegraphics{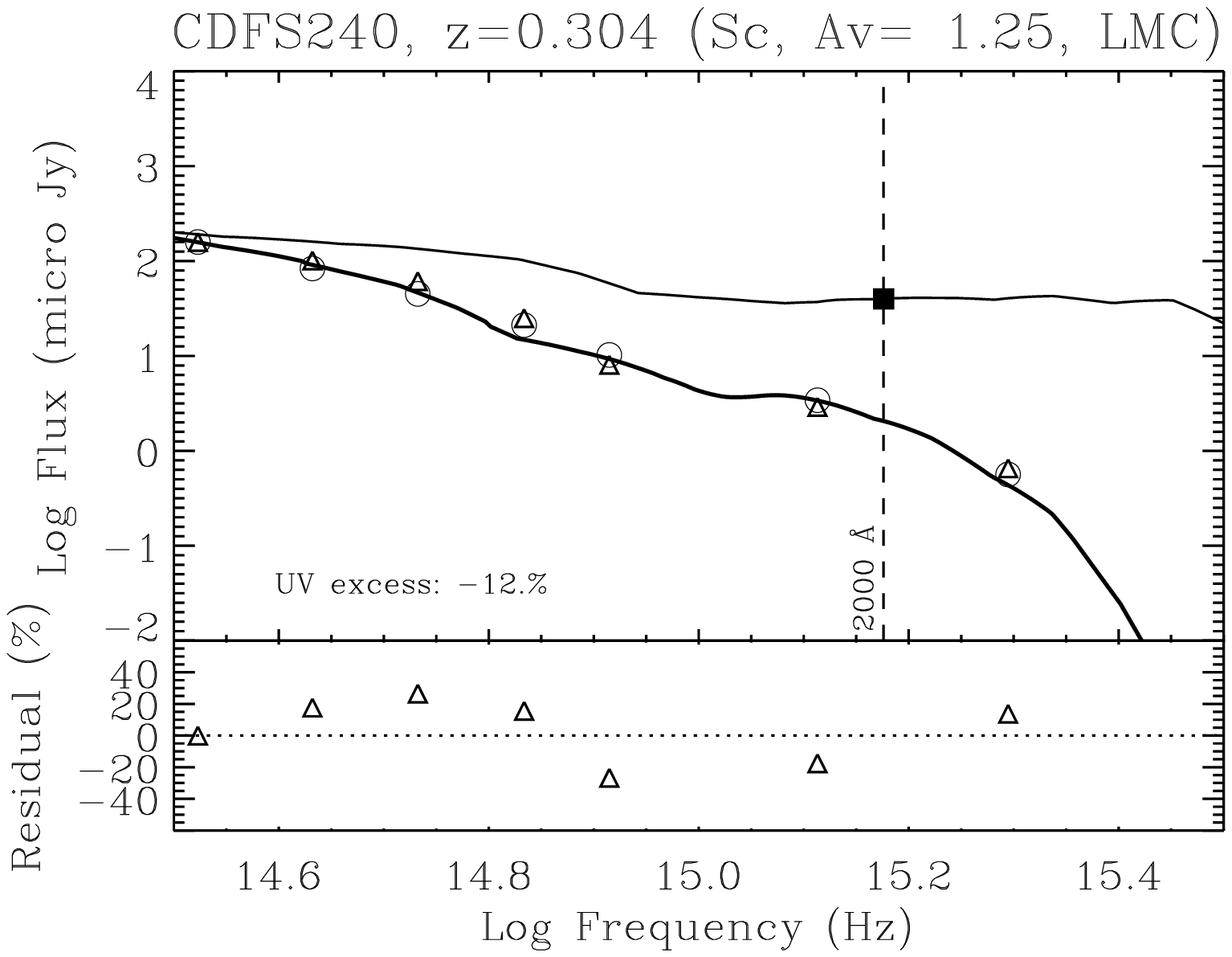}}
\put( 6.5, 4.5){\includegraphics{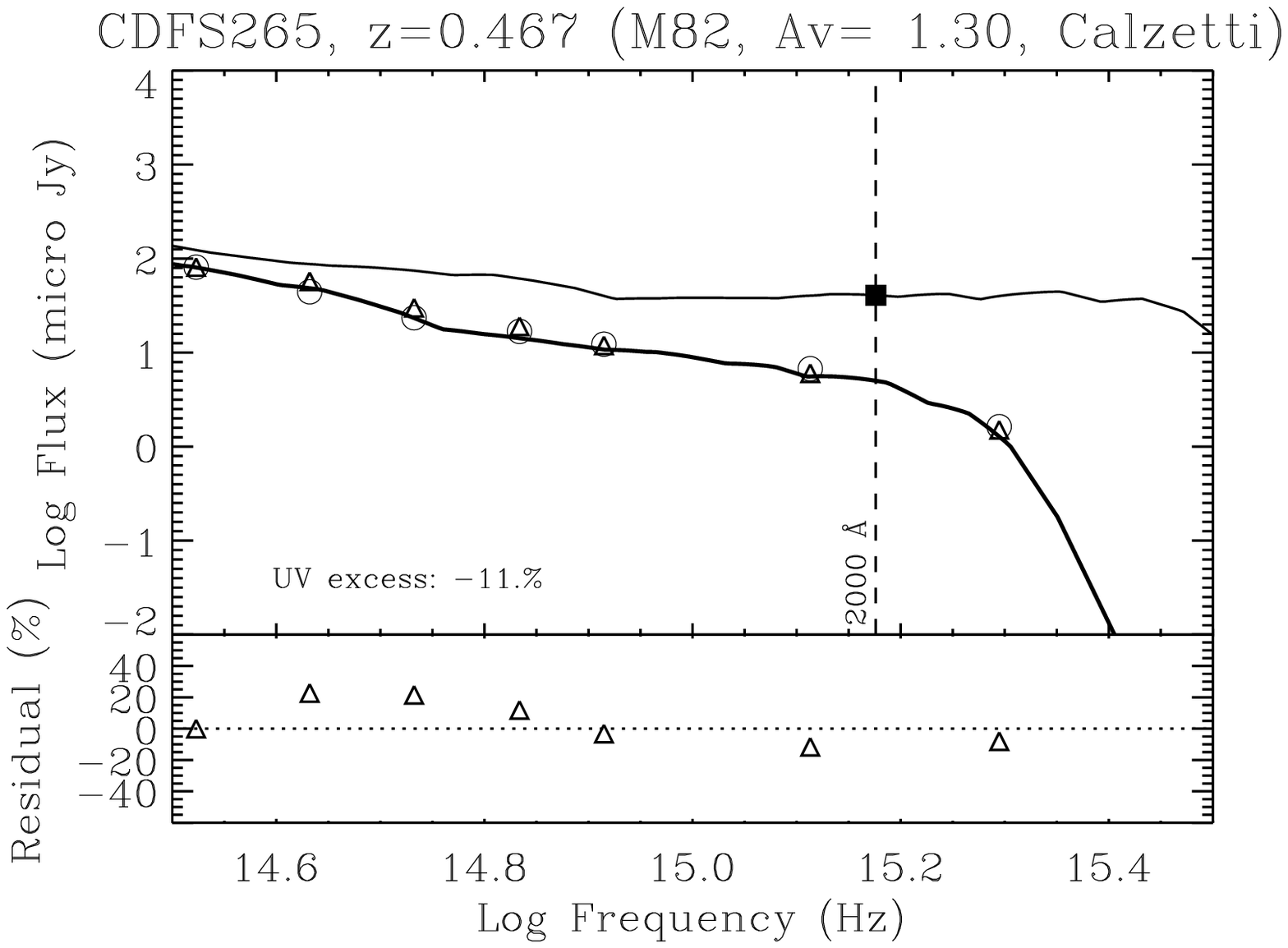}}
\put(-1.5,-2.0){\includegraphics{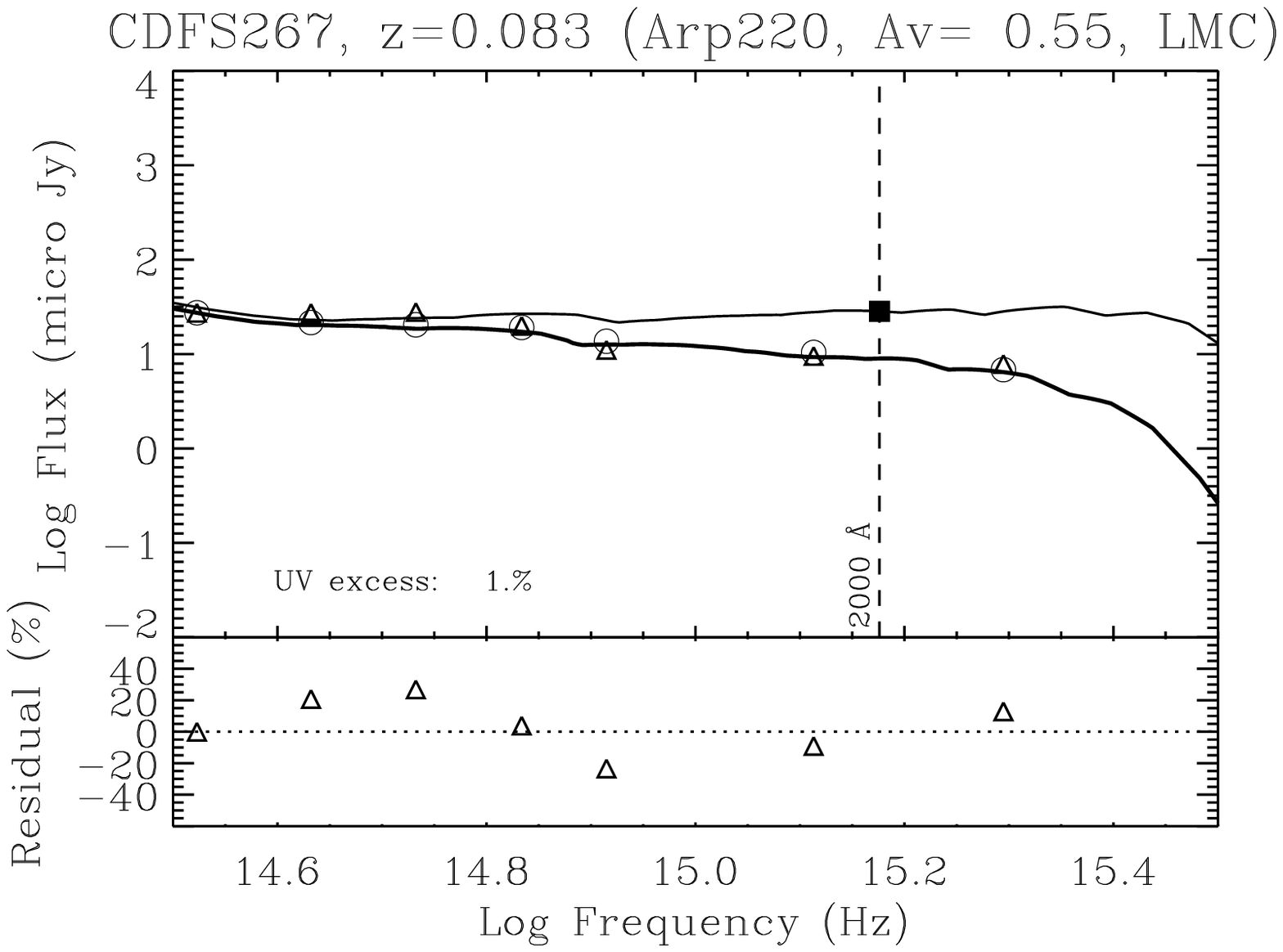}}
\put( 6.5,-2.0){\includegraphics{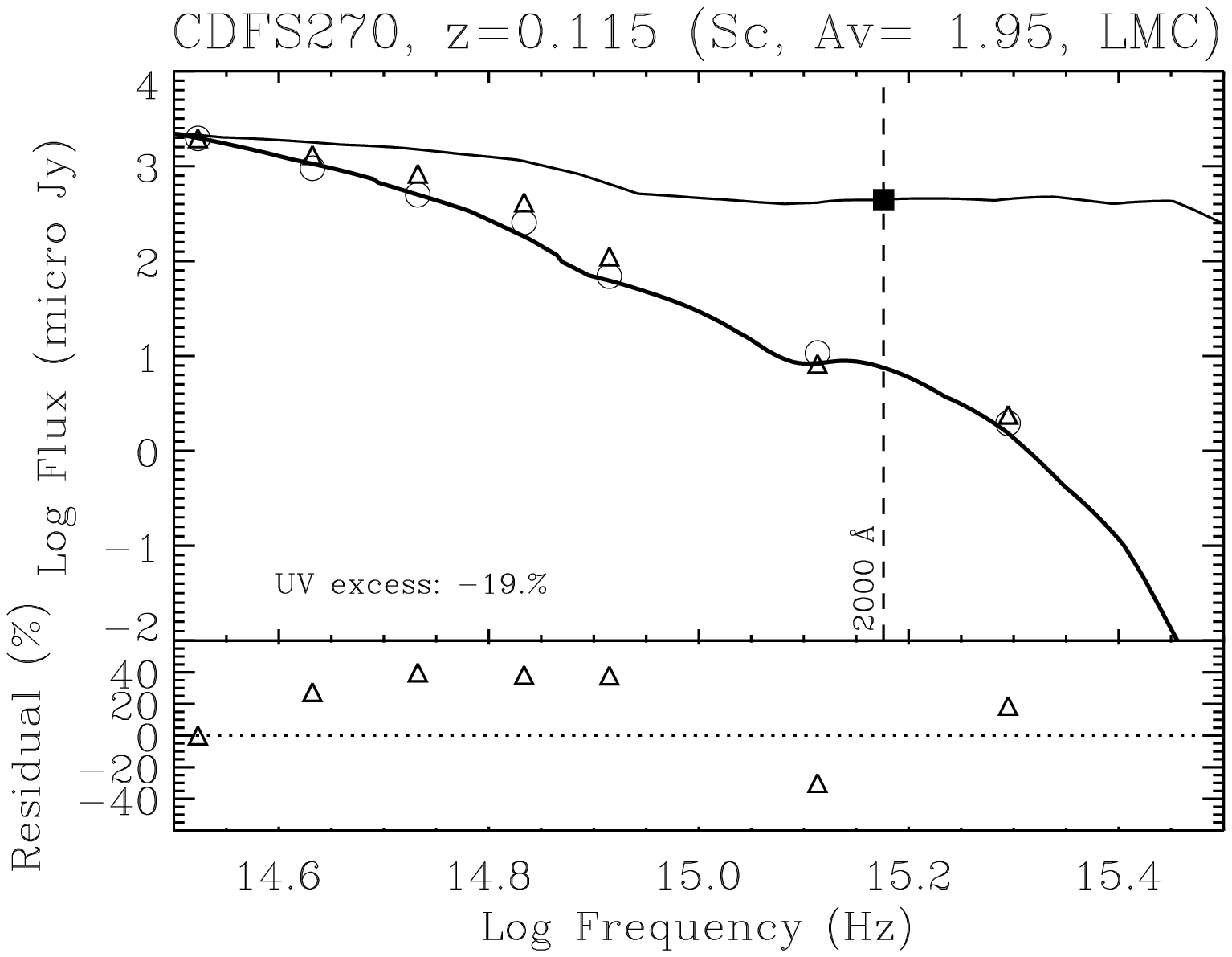}}
\end{picture}
\caption{\label{SEDs_4} 
Continued.
}
\end{figure*}

\setcounter{figure}{0}
\begin{figure*}
\setlength{\unitlength}{1cm}           
\begin{picture}(14,22)       
\put(-1.5,11.){\includegraphics{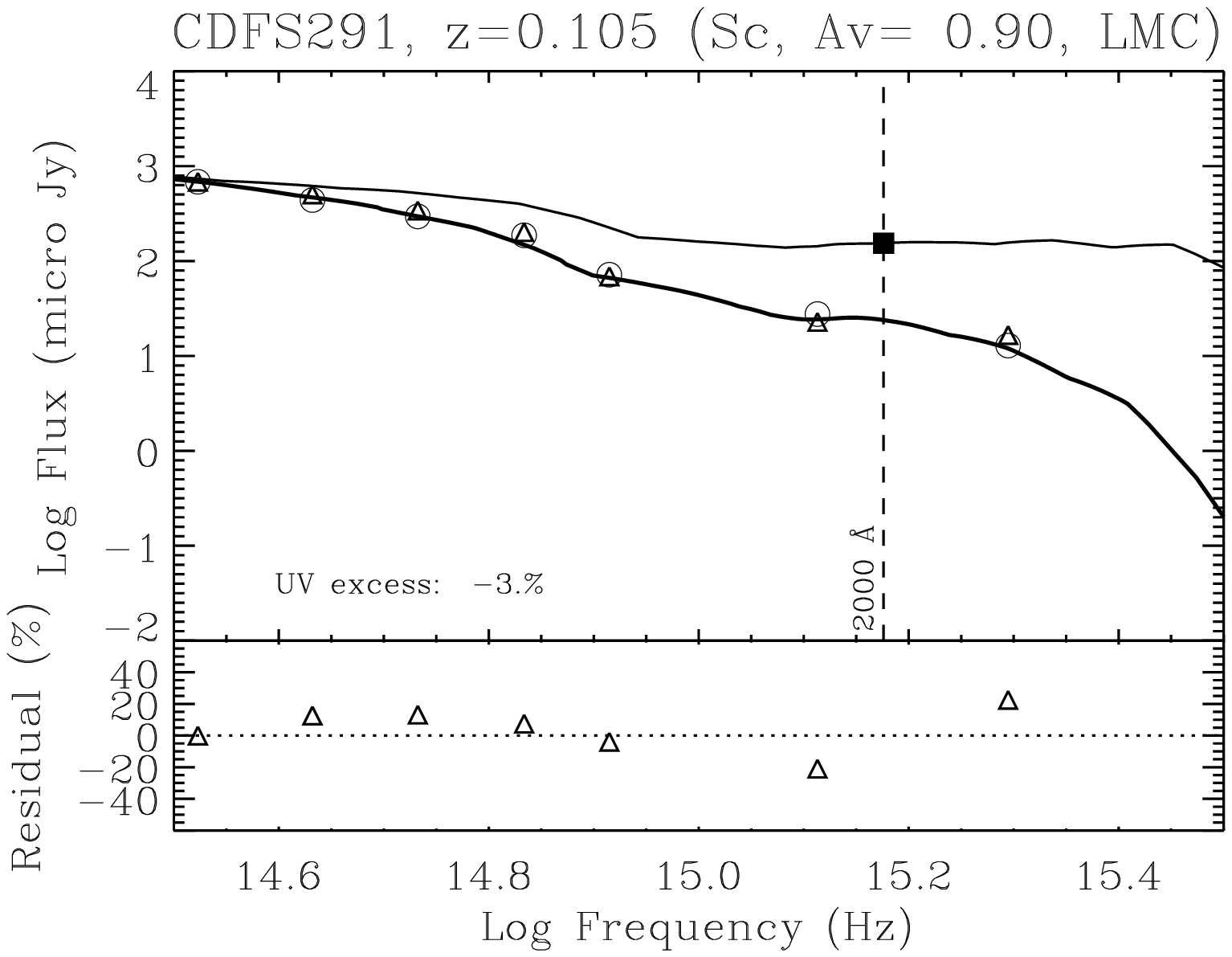}}
\put( 6.5,11.){\includegraphics{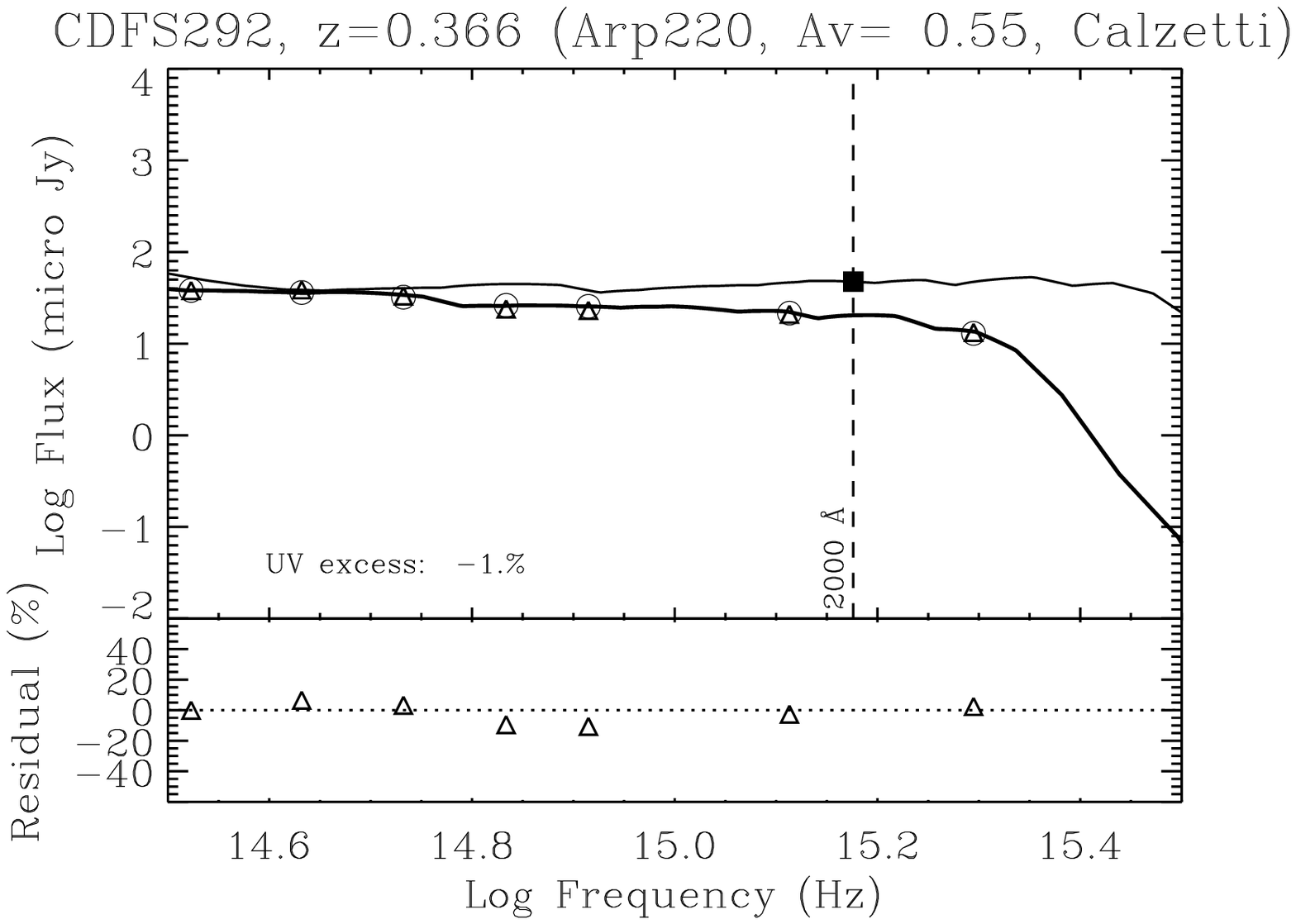}}
\put(-1.5, 4.5){\includegraphics{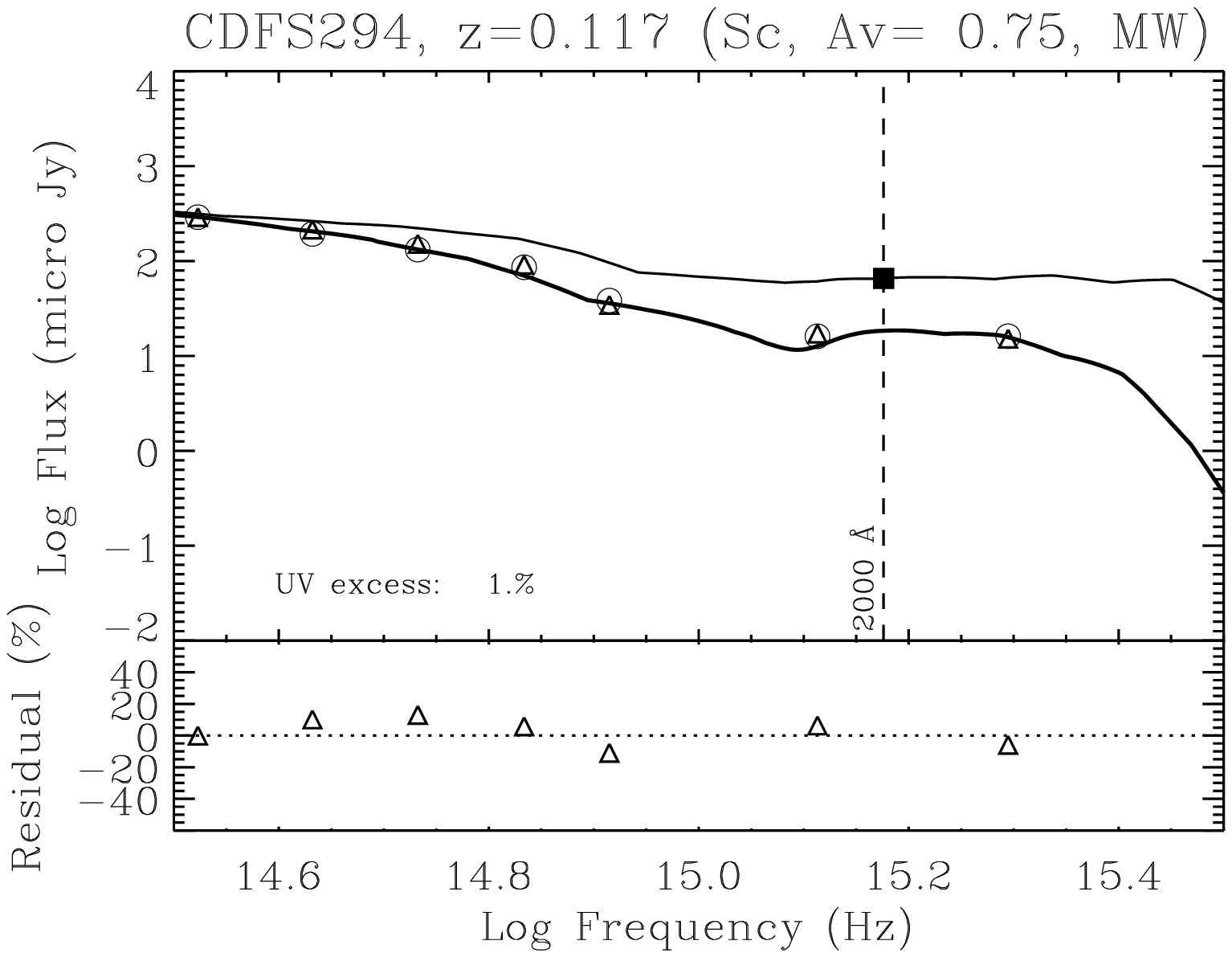}}
\put( 6.5, 4.5){\includegraphics{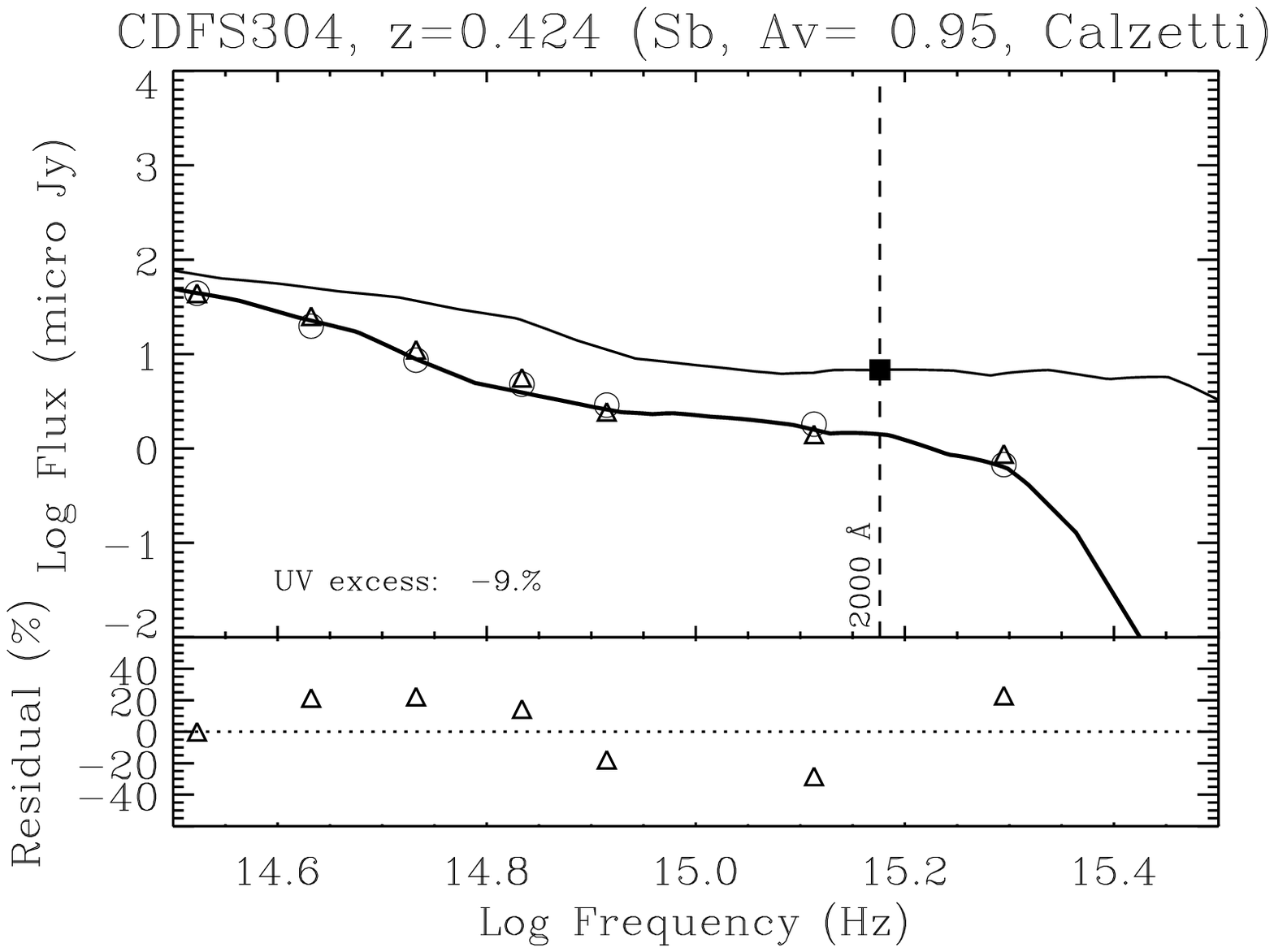}}
\put(-1.5,-2.0){\includegraphics{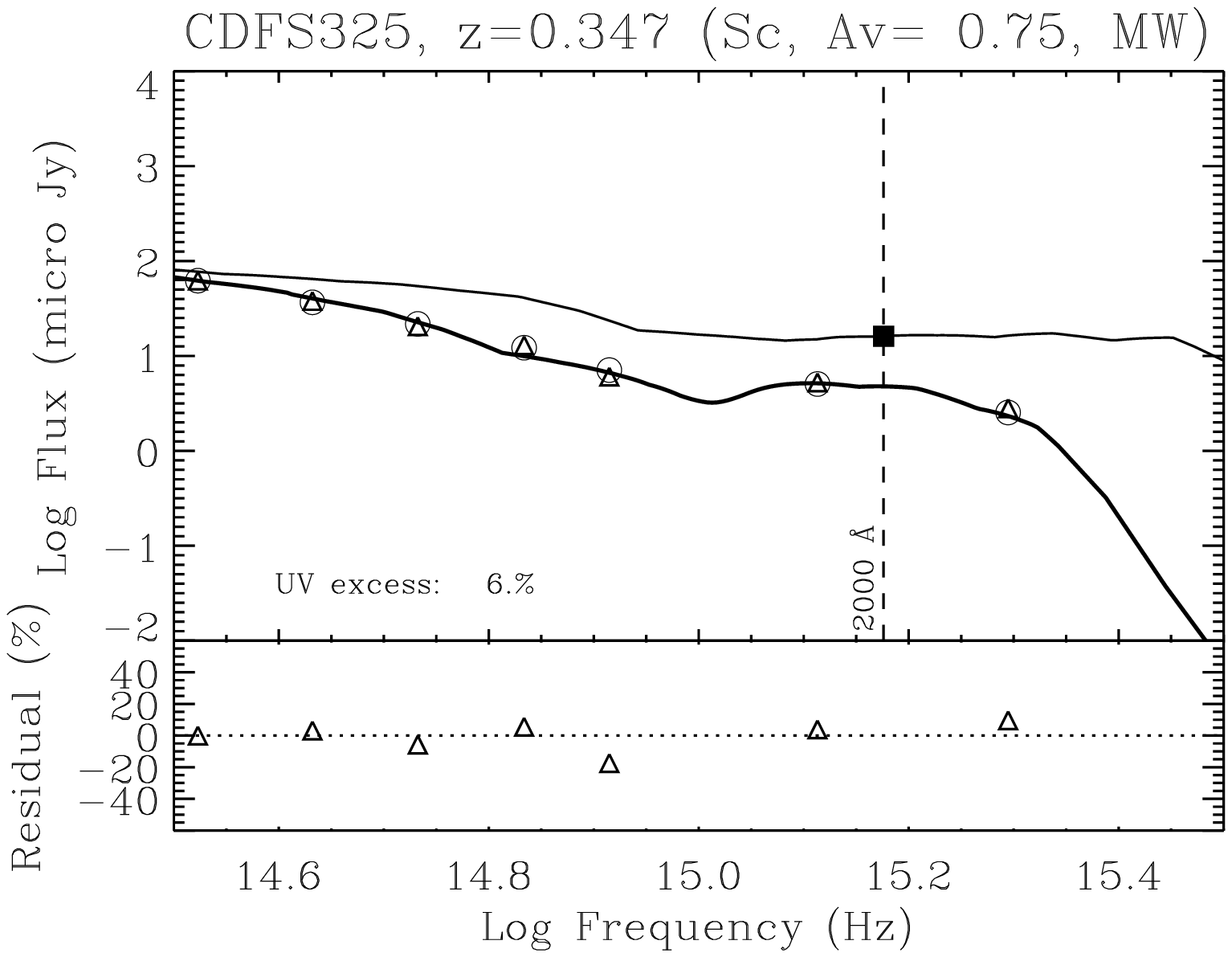}}
\end{picture}
\caption{\label{SEDs_5} 
Continued.
}
\end{figure*}

\section{Correlation between UV and X-ray luminosities}\label{Sec:uvX}

A strong correlation between the star formation rate  calculated from the
UV [SFR(UV)] and  from the X-ray luminositites [SFR(X)]
in the observed galaxies, if it exists, would suggest, na\^\i vely, that these 
luminosities are dominated by a young stellar population. Due to different 
timescales in
   the processes involved, however, there is a very early stage when X-rays 
   would not have yet been produced, as has been discussed in the 
   Introduction.

There are several published relations between the UV luminosity and the 
current star formation rate, that also discuss the difficulty of 
correcting the observed UV fluxes from  extinction  
(e.g. Madau, Pozzetti and Dickinson 1998; 
Rosa-Gonz\'alez, Terlevich and Terlevich 2002).
For solar metallicity and a Salpeter IMF, the SFR(UV)
is given by (Kennicutt 1998), 

\begin{equation}\label{Eq:SFRuv}
SFR(UV) ({\rm M_\odot yr^{-1}})= 1.4\times 10^{-28}  L_{2000} ({\rm erg\, s^{-1}\, Hz^{-1}})
\end{equation}
where L$_{2000}$ is the extinction--corrected 
luminosity per unit of frequency at the rest frame wavelength of 2000\,\AA. 

For the X-ray luminosities we adopted the empirical law 
derived by Ranalli et al. (2003) who combined the 
existing relations between the SFR and the  infrared and radio 
luminosities (Condon 1992, Kennicutt 1998)
with the strong correlation between X-ray and IR 
observed in a sample of local star forming galaxies, 

\begin{equation}\label{Eq:SFRx}
SFR(X) ({\rm M_\odot yr^{-1}}) = 2.2\times 10^{-40} L_{0.2-2 kev}  
\end{equation}

This relation is, within the errors, equal to that by 
Bauer et al. (2002) based on the correlation betweeen 
the SFR obtained by radio and the X-ray luminosities of
galaxies from the Chandra Deep Field North. 
Other calibrations of the SFR(X) are based on the HMXB 
luminosities (e.g. Grimm et al. 2003, Persic et al. 2004). The relation 
between the integrated luminosity of the HMXBs in the 2--10 keV band 
strongly correlates with the SFR given by the IR luminosities 
and shows a very small scatter, smaller in fact
than the scatter observed if the total 
X-ray luminosity is used.  However due to the impossibility of
separating the HMXB from the total luminosity we are forced to use
Ranalli et al. (2003) empirical relation.

\begin{figure}
\setlength{\unitlength}{1cm}           
\begin{picture}(7,9)       
\put(-1.5,-2.5){\includegraphics{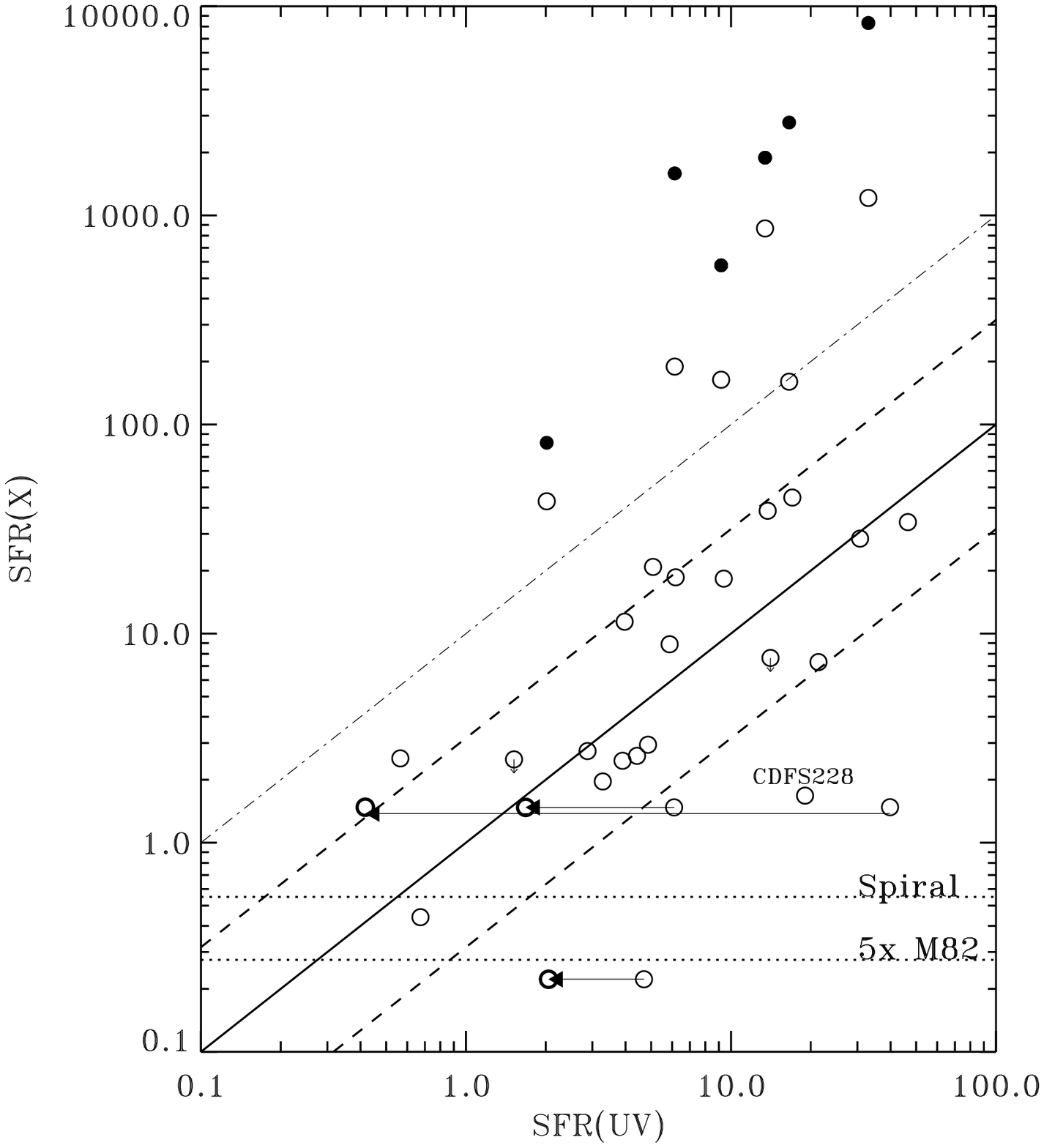}}
\end{picture}
\caption{\label{SFR-uvX} 
Open circles represent the 
SFR given by the UV versus the SFR given by the soft X-ray [SFR$^{soft}$(X)].
The solid line represents equal values and the dashed lines a 
deviation of 0.5 dex. 
Solid circles are the SFR calculated using the hard X-ray  [SFR$^{hard}$(X)] 
for the  
galaxies with $\Delta$XUV greater than 1 (dot--dashed line).
The arrows 
point to the SFR(UV) calculated using the empirical relation 
between the FIR luminosity and the extinction in the UV 
for galaxies with $\Delta$XUV $\le$-0.5 detected by Spitzer. CDFS228, the 
outlier not observed by Spitzer, is marked.   
The horizontal dotted lines show the contamination 
due to LMXB for a galaxy with 5 times more stars than M82 and for a typical 
spiral galaxy, as labelled (see text).
}
\end{figure}

Figure~\ref{SFR-uvX} shows the SFR(X) obtained from equation~\ref{Eq:SFRx} 
against SFR(UV).  We find that most of the galaxies lie close to the 
solid line defined as SFR(X)=SFR(UV) 
even at low values of the SFR(UV) ($\la~5$~\Msolar~yr$^{-1}$); 
   this contradicts previous claims by Grimm, Gilfanov and Sunyaev (2003)
   which are based on the luminosity of HMXB alone.
   In contrast, we would like to remark that the
   expression (used by us) given by Ranalli et al. (Equation~\ref{Eq:SFRx})
   is  obtained by direct comparison between radio and IR luminosities and
   therefore, it refers to the {\bf total} X-ray luminosity 
which includes, apart from the contribution due to 
HMXB and LMXB, the emission from young SNRs and the diffuse hot gas.
Notice also that when the X-ray 
luminosity is low, the SFR(X) derived from HMXB emission could be severely
affected by stochastic effects since the flux would be dominated by a few
binaries. 
The strong correlation found by Ranalli et al. extends to 
low values of the SFR ($\la$~0.1 \Msolar\ yr$^{-1}$), 
and the presence of LMXB that are not associated with the 
present star formation activity seems to be a minor effect.  
However, for galaxies with SFR smaller than  1\Msolar year$^{-1}$\ and 
masses typical of a spiral galaxy, the X-ray luminosities 
and the derived SFR(X) could have an important contribution from LMXB
affecting the observed correlation. 
We estimate in what follows the X-ray luminosity due to LMXB.
Grimm, Gilfanov and Sunyaev (2002) found a relation between 
the stellar mass and the total X-ray luminosity produced by LMXBs in the 
Galaxy.
They compared the integrated luminosity function of galactic 
LMXB detected by the All-Sky Monitor aboard the Rossi X-ray 
Timing Explorer  and assumed a stellar mass for the Galaxy of 
5$\times 10^{10}$\Msolar, in order to get the X-ray luminosity per unit 
stellar mass due to LMXB, 

\begin{equation}\label{eq:LMXB}
L_X (LMXB) ({\rm erg\, s^{-1}}) = 5\times 10^{28}  M_s 
\end{equation}
where the stellar mass ($M_s$) is given in solar masses.

    By combining the derived stellar masses (Table~\ref{Tab:CDFS-Templates}) 
    with equation~\ref{eq:LMXB} 
    we computed the luminosity produced by the LMXB and the percentage 
    of the derived SFR(X) not related to the recent star formation activity.
    The obtained values presented in Table~\ref{Tab:CDFS-Templates} show that 
    the contamination of the estimated SFR(X) due to LMXB 
    is lower than 20\% for all the cases except for CDFS213.
    In fact the X-ray luminosity of CDFS213 could be fully explained as due to
    emission from 
    LMXBs. This ``extreme" galaxy is further discussed in Section~\ref{AGN}.
    In order to visualize the LMXB contribution to the SFR(X) we calculated
    the X-ray luminosity for a starforming galaxy having a stellar mass 
    somewhere between the
    prototype starburst galaxy (M82;  stellar mass 
    $M_s = 5\times 10^9$\Msolar ,  Mayya et al. 2006)
    and the typical mass of a spiral galaxy like the Milky Way 
    (MW, $M_s = 5\times 10^{10}$\Msolar , Grimm et al. 2002). 
    We took arbitrarily $M_s$ as 5 times larger than that
    of M82 and represented the corresponding SFR(X) in Figure~\ref{SFR-uvX}
    by a dotted line.

\begin{figure}
\setlength{\unitlength}{1cm}           
\begin{picture}(7,9)       
\put(-1.5,-2.5){\includegraphics{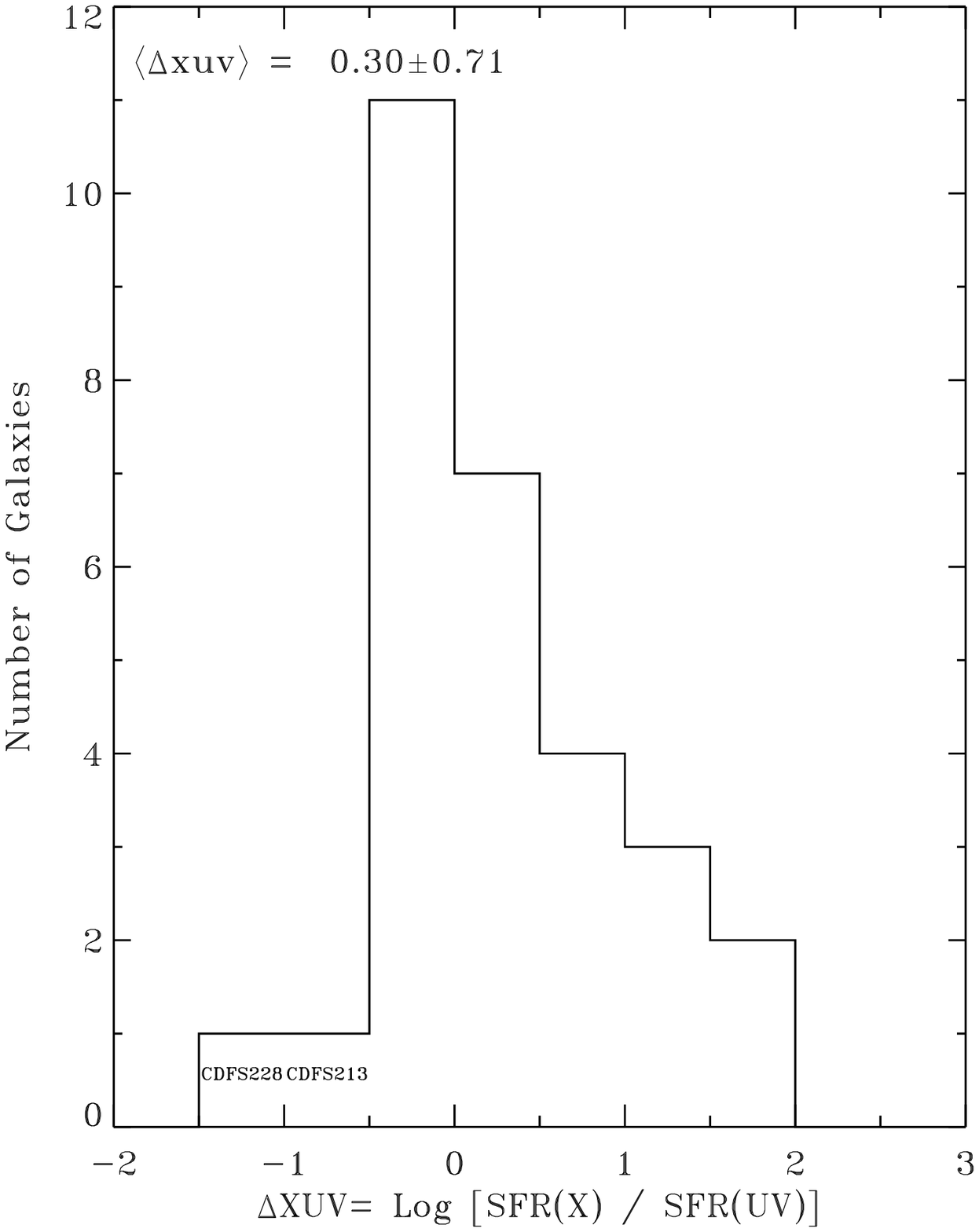}}
\end{picture}
\caption{\label{HISTuvX}
Histogram of the relation between the SFR given by the UV and
by the X-rays. The mean and standard deviation are shown at the 
top. 
}
\end{figure}

     Figure~\ref{HISTuvX} shows the histogram of the logarithm of the 
     ratio between the SFR(X) and SFR(UV)  ($\Delta$XUV). 
     The histogram shows clearly that for the majority of the galaxies 
     the SFR(X) is similar to SFR(UV), and that there are five galaxies  
     with  $\Delta$XUV $>$1, and one galaxy with  $\Delta$XUV$<-1$.
     The nature of these galaxies is discussed in the next section. 
      
\section{Contamination by Hidden AGN and the Discovery of
X-ray Weak Galaxies}
\label{AGN}

Figures~\ref{SFR-uvX} and~\ref{HISTuvX} show that most of 
the galaxies ($\sim$62\% of the sample) lie within $\pm$0.5 dex of the line 
which defines SFR(UV) = SFR(X) ($\Delta$XUV=0)
consistent with them being normal star forming galaxies. 
However, when  SFR(UV) is
higher than 5\Myear\ the scatter is significantly larger.  
Similar amounts of scatter reaching lower SFRs are found in other samples of 
galaxies when Seyfert 2 and LINER galaxies are included in the 
relation between the SFR and the X-ray luminosity (e.g. Ranalli et al. 2003).

\begin{figure*}
\setlength{\unitlength}{1cm}           
\begin{picture}(7,6)         
\put(-6.5,-1.5){\includegraphics{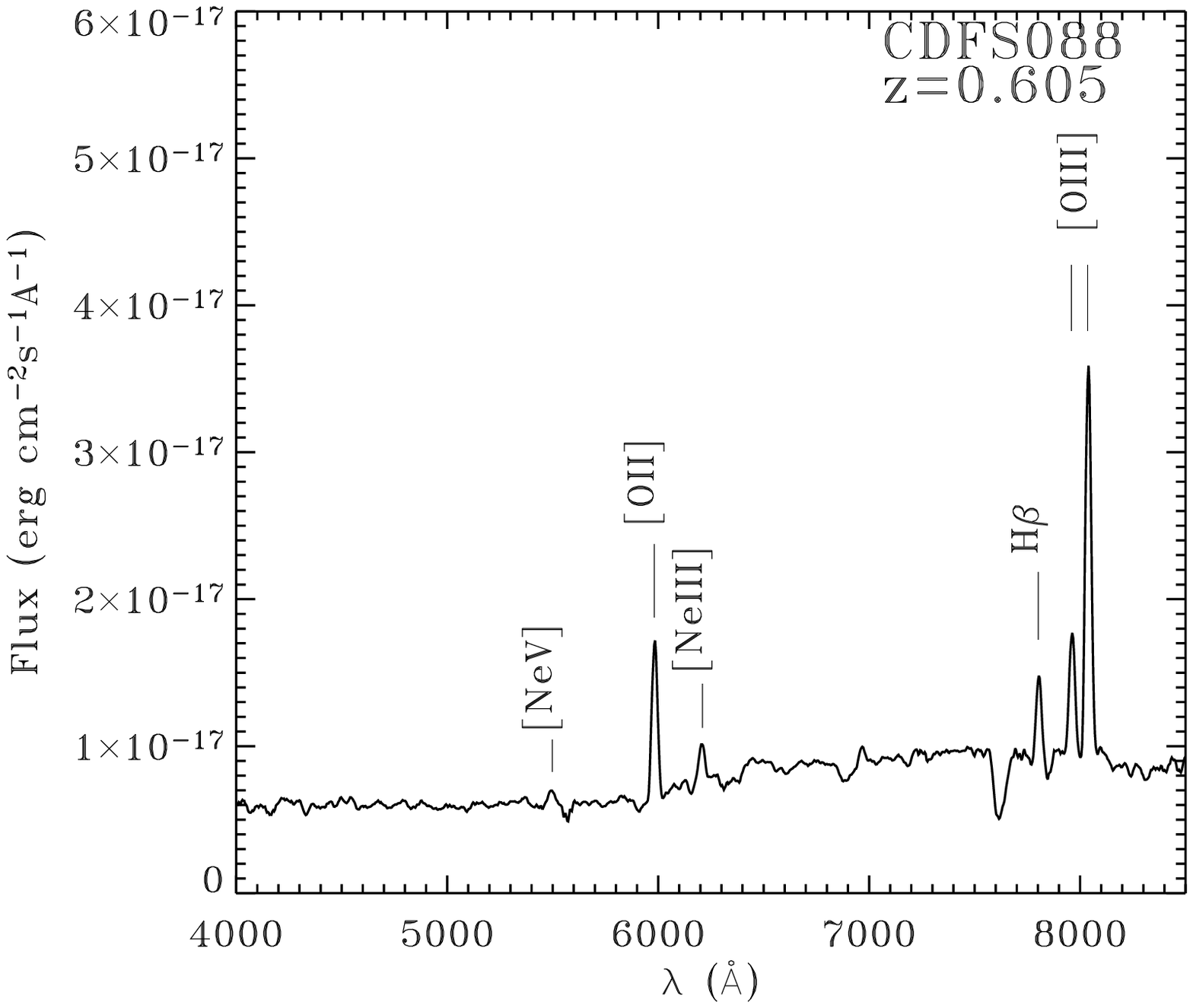}}
\put(-0.5,-1.5){\includegraphics{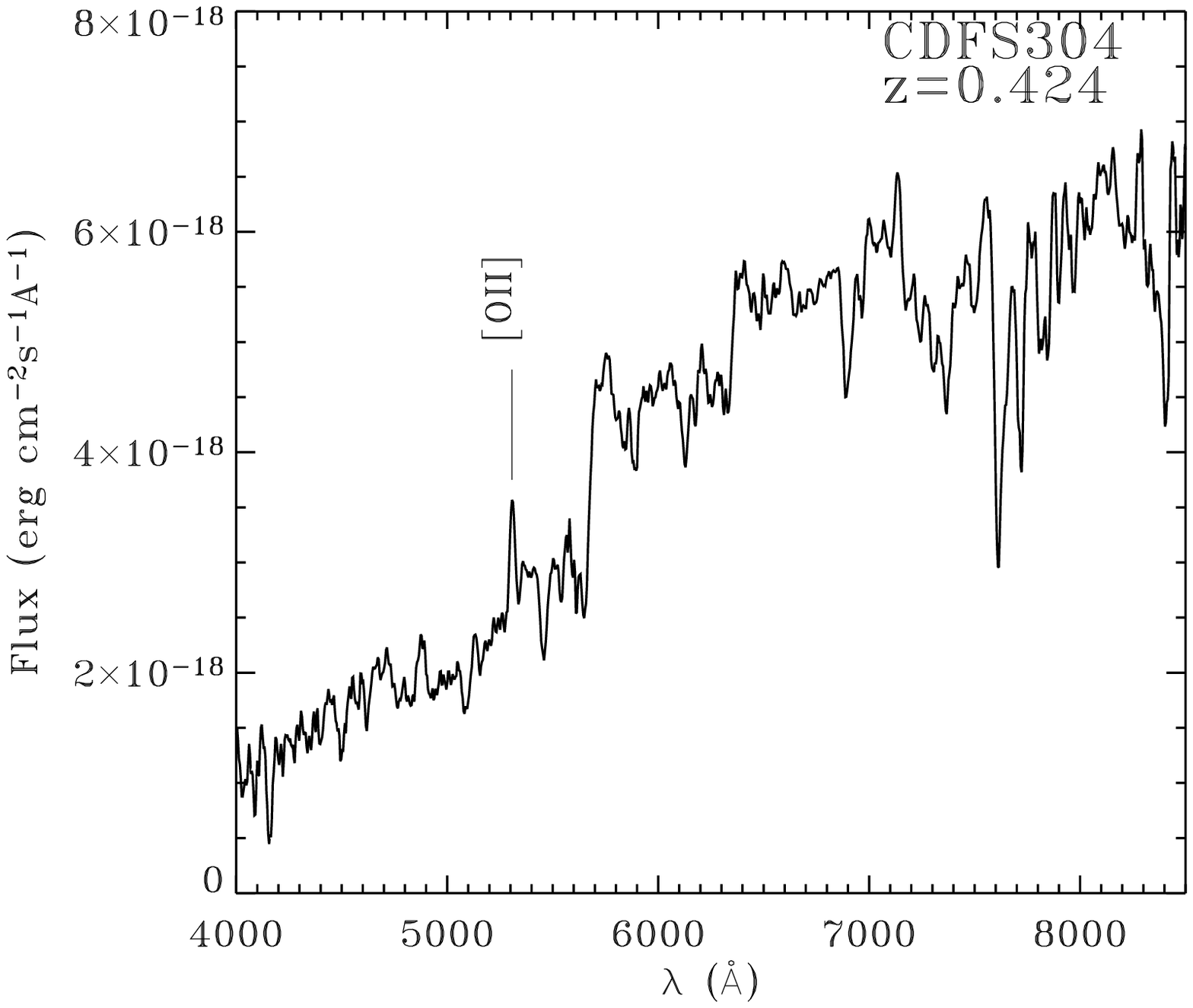}}
\put(5.8,-1.5){\includegraphics{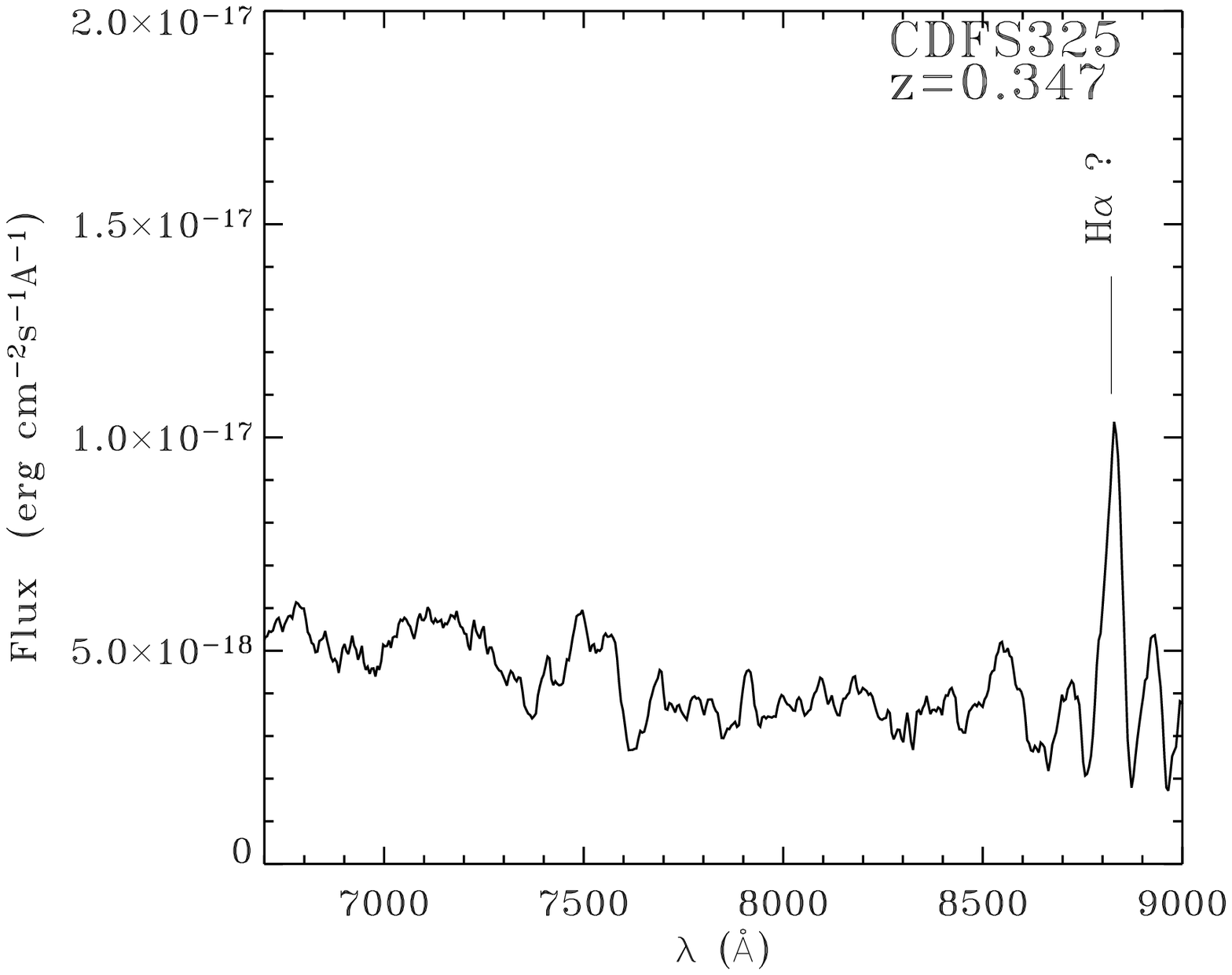}}
\end{picture}
\caption{\label{Szokoly} VLT spectra of three of the five 
galaxies with $\Delta$XUV $>$ 1. The spectra were extracted 
from Szokoly et al. (2004).}
\end{figure*}

The large scatter seems to be  related to two different phenomena: 
i) galaxies that could be contaminated by the presence of 
very obscured AGN that will cause the X-rays to be bright and the UV, faint, 
and ii) X-ray weak galaxies due to a small number of HMXB for a given SFR,
or to an overcorrection of the observed UV fluxes.
\noindent i) All  galaxies with SFR(X) higher than 100~\Myear\ 
show a $\Delta$XUV larger than 1; they 
could be contaminated by a very obscured AGN not 
present in the optical SED but which nevertheless is powering the X-rays 
(Comastri et al. 2001, Maiolino et al. 2003 and references therein).
Figure~\ref{Szokoly} shows VLT spectra 
of three of these galaxies for which 
$\Delta$XUV is larger than 1. 
The spectra were  extracted from  Szokoly et al. (2004)
\footnote{
Szokoly et al.~spectra obtained from the dedicated page: 
http://www.mpe.mpg.de/CDFS/data/}.

CDFS088 is the only one of the three galaxies that shows high 
excitation emission
lines (e.g. [Ne V]) and strong oxygen lines suggestive of 
nuclear activity. The other two galaxies do not present strong emission lines. 
They are classified as normal galaxies (note that Szokoly et al. 2004 
classified these two galaxies as low emission line galaxies). We would like to 
remark, however, that strong optical lines due to the presence of an AGN could
have fallen outside the spectral optical range.

Hard X-rays may indicate that an AGN is
present and this can be tested by calculating the SFR given 
by hard X-rays [$SFR^{hard}(X)$],  and comparing
it to the SFR(X) given by the soft band.

All the galaxies with SFR(X) higher than 100~\Myear\  
were detected in the hard X-ray band (Table~\ref{Tab:CDFS-X-uv} and
Table~\ref{Tab:CDFS-Templates}).
The corresponding SFR was calculated as (Ranalli et al. 2003), 
\begin{equation}\label{Eq:SFRxHard}
SFR^{hard}(X) ({\rm M_\odot yr^{-1}}) = 2.\times 10^{-40} L_{2-10 kev} (\rm erg\, s^{-1})
\end{equation}
The calculated $SFR^{hard}(X)$ in these galaxies is higher than that 
obtained from the soft X-rays (see solid circles in Figure~\ref{SFR-uvX}), 
supporting the presence of an obscured AGN (or partially obscured as for 
CDFS088).

The given SFR(X) in the soft and hard bands are based on the empirical 
results from Ranalli et al. (2003) which compared the X-ray with 
IR and radio luminosities closely related to recent star forming events. 
The observed scatter is around 0.3 dex covering the  X-ray luminosity range 
from  10$^{38}$ to 10$^{42}$ erg s$^{-1}$. If the 
observed luminosities are due to LMXB, an X-ray luminosity of
4.5$\times$10$^{41}$ erg s$^{-1}$ (equivalent to a SFR of 100\Msolar\ 
year$^{-1}$) would imply the existence of 
a galaxy with an unfeasible stellar mass of about 10$^{13}$\Msolar.
In all the galaxies with SFR(X) greater than 100 \Msolar\ year$^{-1}$
and $\Delta$XUV higher than 1 the estimated contamination due to LMXB 
is lower than 1\% (Table~\ref{Tab:CDFS-Templates}). The suggestion being  
therefore that most of the X-ray luminosity for these objects is 
related to the presence of a
central massive black hole, and the contribution due to 
LMXB and star forming processes is quite small.

CDFS078 is one of the galaxies for which $\Delta$XUV is higher than 1.
In this case the SFR(UV) is about 2 \Msolar\ year$^{-1}$ but 
the SFR(X) is 43 and the SFR$^{hard}$(X) is close to 
100. Unfortunately this galaxy has no optical spectrum but it seems
to be a low luminosity counterpart of the possible active galaxies discussed 
above.

Notice that for galaxies with $\Delta$XUV$>$1, the absolute value of the 
UV excess defined 
as the difference between the observed UV flux and the flux given by the model 
is below 20\%, therefore, the model seems 
to be a good representation of the 
stellar content of the galaxy and the observed UV fluxes are not heavily 
affected by the presence of an AGN.

In any case, we can not rule out the presence of an obscured,  compact and 
very powerful  star forming region of which we are only detecting 
a small fraction of the  produced UV-optical emission.  
This effect known as {\it age selective extinction} 
is observed  in the central regions of  
powerful starburst galaxies  (e.g. Mayya et al. 2004).

\noindent ii) Two galaxies CDFS207 and CDFS228 with SFR(UV) higher 
than 15\Myear\ show very low X-ray emission implying 
SFR(X) of about 1.5~\Myear. 
These galaxies are the most obscured ones in our sample 
(A$_V$=2.85 and 2.35 respectively)
and are compatible with an Arp~220-like template. 
Of them, only CDFS228 was detected in the hard X-rays band, and the deduced
SFR$^{hard}(X)\sim 7$  is still 6 times lower than the 
corresponding SFR(UV). 

For Arp220, the ratio between the X-ray luminosities and the UV corrected by
extinction
L$_X/(\nu$ L$_{UV}$)= 1.7$\times 10^{-3}$ (UV data obtained 
from Goldader et al. 2002 and X-ray data from Iwasawa et al. 2001), 
is similar to the values obtained for CDFS185 and CDFS207 
of 1.2$\times 10^{-3}$ and 1.6$\times 10^{-3}$ respectively. 
However CDFS228 has a value of L$_X/(\nu$ L$_{UV}$)= 4$\times 10^{-5}$ 
indicating that we are probably overestimating the UV corrections.
Unfortunately this galaxy was not observed by Spitzer so 
we could not further analyse the  correction factor 
applied to the UV fluxes as discussed in the following section.
The ratio L$_X/(\nu$ L$_{UV}$) for the other two galaxies fitted with 
an Arp220-like spectrum 
is about a factor of four below the Arp220 value. 

Another extreme galaxy is CDFS213 where 
an important fraction of the observed X-ray flux 
could be due to the presence of LMXBs (see Figure~\ref{SFR-uvX}).
In any case, even if the observed X-ray flux has a 
strong contribution from LMXBs which are not related to a recent 
burst, the estimated $\Delta$XUV is smaller than -1. 
For estimating the SFR(X) in this galaxy, we are using the 
flux provided by Giacconi et al. (2002) instead of the upper limit 
provided by Alexander et al. (2003). If we consider that
the observed X-ray luminosity is just an upper limit, the 
nature of this galaxy would be even more extreme. 

X-ray emission in the youngest objects is dominated by HMXB, therefore
a delay between the stellar UV-continuum from
massive stars (proportional to SFR(UV))  and the first compact  object that 
is created after a supernova event (responsible for 
the deduced SFR(X)) is expected. 
A similar delay between the formation of stellar clusters and the
production of compact objects after supernova explosions is observed in 
radio. 
In fact,  the lack of synchrotron radiation has helped to identify
young bursts of less than $\sim$20 Myrs 
(Rosa-Gonz\'alez et al. 2007).

There is always the possibility that we are just
overestimating the correction factors applied to the UV fluxes and therefore
we are not observing young objects.
It has been indicated that in some
cases the fluxes in the UV range could be badly estimated 
(e.g Burgarella, Buat, \& Iglesias-P{\'a}ramo 2005, Buat et al. 2005). 
We explore this possibility in the next section using Spitzer 
observations  of the CDFS.

\section{Analysis of Spitzer Data}
\label{Spitzer}

The release (DR3) of the Great Observatories Origins Deep Survey (GOODS), 
one of Spitzer's Legacy Science Programs, includes  MIPS 
\footnote{The Multiband Imaging Photometer for Spitzer (MIPS) is fully
  described in Rieke et al. (2004).} 
24$\mu$m images for GOODS-S and the corresponding source list
~\footnote{http://data.spitzer.caltech.edu/popular/goods/Documents/\linebreak[4]goods\_ dataproducts.html}. 
Table~\ref{Tab:CDFS-Spitzer} shows the fluxes at 24 $\mu$m (S$_{24}$) obtained 
by Spitzer for our sample of CDFS galaxies. 
Notice that 5 galaxies were outside the Spitzer coverage.
The Spitzer fluxes are from DR3 except for galaxies
with a Spitzer index of 000 (see Table~\ref{Tab:CDFS-Spitzer}).
The fluxes of these galaxies were extracted directly from the scientific 
maps,  using the conversion factor between observed counts and
fluxes provided by the GOODS team (e.g. Dickinson \&  GOODS 2004).

To estimate the total infrared luminosities (integrated between 8 and 
1000$\mu$m) the 24$\mu$m luminosities obtained from the observed fluxes were 
transformed to infrared luminosities using
the empirical relation given by Takeuchi et al. (2005) 

\begin{equation}\label{eq:Takeuchi}
log\, L_{IR} = 2.01 + 0.878\, log\, L(25 \mu m)
\end{equation}
where both luminosities are given in solar units.
This relation is based on the study of 1420 galaxies with available 
data in the four IRAS bands. Takeuchi et al. (2005) show  
that this formulation provides good estimates of the total IR flux in a 
wide variety of galaxies including  extreme galaxies such as Arp220.

The derivation of
the $L(25 \mu$m) from the observed S$_{24}$ fluxes, requires to
assume a spectral energy distribution and to estimate 
the corresponding $K$--correction.
For each galaxy we used the template obtained by the optical fit 
described in Section~\ref{Uvfluxes} but instead of using templates 
that only contain stellar light we use those with dust emission included.
A full description of the templates and the application to 
model galaxies is given in 
several papers by the GRASIL group 
(e.g. Silva et al. 1998, Panuzzo et al. 2005, Vega et al. 2005).
The luminosity at a given frequency ($\nu$) is given by 
\begin{equation}
L_\nu=  4\pi D_L^2   \frac{S_{\nu o}}{(1+z)} 
\end{equation}
where $D_L$ is the luminosity distance, $S_{\nu o}$ is the observed flux at 
24$\mu$m, $z$ is the redshift of the source and 
$\nu= \nu o (1+z)$.
Once we get $L_\nu$ we normalize the corresponding template 
to this value and find the luminosity at $\lambda= 25\mu$m.


\subsection{Using the FIR to correct Dust Attenuation}

We have explored the possibility that  
galaxies with $\Delta$XUV $\le$-0.5 are outliers 
in Figure~\ref{SFR-uvX} because the estimated dust attenuation is biased or 
not accurate enough in the absence of FIR constraints. 

To this end we
computed from the observed FIR a new value for the corrected  flux at 
2000 \AA\ (F2000$^*$), combining the IR luminosities obtained in the 
previous section with the GALEX data.
Following  Burgarella et al. (2005), the extinction in the 
GALEX bands (A$_{FUV}$ and A$_{NUV}$) is related to the IR flux 
(F$_{\rm IR}$) by 

\begin{eqnarray}\label{eq:NewAv}
A_{FUV} = &-0.028 [Log (F_{IR}/F_{FUV})]^3 \\
          &+ 0.392 [Log (F_{IR}/F_{FUV})]^2 \nonumber \\
          &+ 1.094 [Log (F_{IR}/F_{FUV})] + 0.546 \nonumber \\
           &      \nonumber\\
A_{NUV} = &-0.075 [Log (F_{IR}/F_{NUV})]^3  \nonumber\\
 &+ 0.639 [Log (F_{IR}/F_{NUV})]^2  \nonumber\\
 &+ 0.673 [Log (F_{IR}/F_{NUV})] + 0.260 \nonumber
\end{eqnarray}
where F$_{FUV}$ and  F$_{NUV}$ are the rest frame
far-UV and near-UV fluxes respectively. 
F$_{FUV}$ and  F$_{NUV}$, together with the rest frame F2000,  
were measured from the fitted spectrum 
obtained in section~\ref{Uvfluxes}. 
Relation (8) was used to obtain the 
extinction in the far- and near- UV and 
the extinction at 2000 \AA\ was calculated as
the mean value of $A_{FUV}$ and $A_{NUV}$.
The extinction corrected F2000 (F2000$^*$) was used to
estimate the corrected SFR(UV). 

The  results of this correction are shown in Figure~\ref{SFR-uvX}, 
where the  arrows connect the original SFR(UV) with the new corrected values. 
In two cases (CDFS207 and CDFS291) the new
walues are much closer to the SFR(X), but for
CDFS213 the new SFR(UV) is still below 
the SFR given by the X-ray luminosities by a factor of about 10.
This difference between the SFR estimators will remain even after
taking into account that CDFS213 may be
affected by the presence of LMXBs as showed in
Figure~\ref{SFR-uvX} and discussed in section~\ref{Sec:uvX}.

Unfortunately CDFS228 (labelled in the figure)
is outside the Spitzer field of view  and 
it was not possible to perform for it the analysis just discussed.

\subsection{Infrared SFR}
\label{SEC:SFR(IR)}

A significant fraction of the optical and UV light emitted by young and 
massive stars is 
absorbed by dust grains and re-emitted in the infrared regime  
making this wavelength  a good tracer of the current star formation 
activity as confirmed by multiple observations (e.g. Yun, Reddy \& 
Condon 2001, Schmitt et al. 2006). A relation between the IR and the 
current star formation rate for normal galaxies is given by Kennicutt (1998) 
\begin{equation}\label{eq:SFR_IR}
SFR(IR) ({\rm M_\odot yr^{-1}}) = 4.5 \times 10^{-44} L_{IR} (erg\, s^{-1})
\end{equation}

Notice that $L_{IR}$  refers to the luminosity integrated over the 
full IR espectrum covering from 8 to 1000 $\mu$m, therefore it is equivalent 
to the L$_{IR}$ calculated by equation 6. 

\begin{figure}
\setlength{\unitlength}{1cm}           
\begin{picture}(7,9)       
\put(-1.5,-2.5){\includegraphics{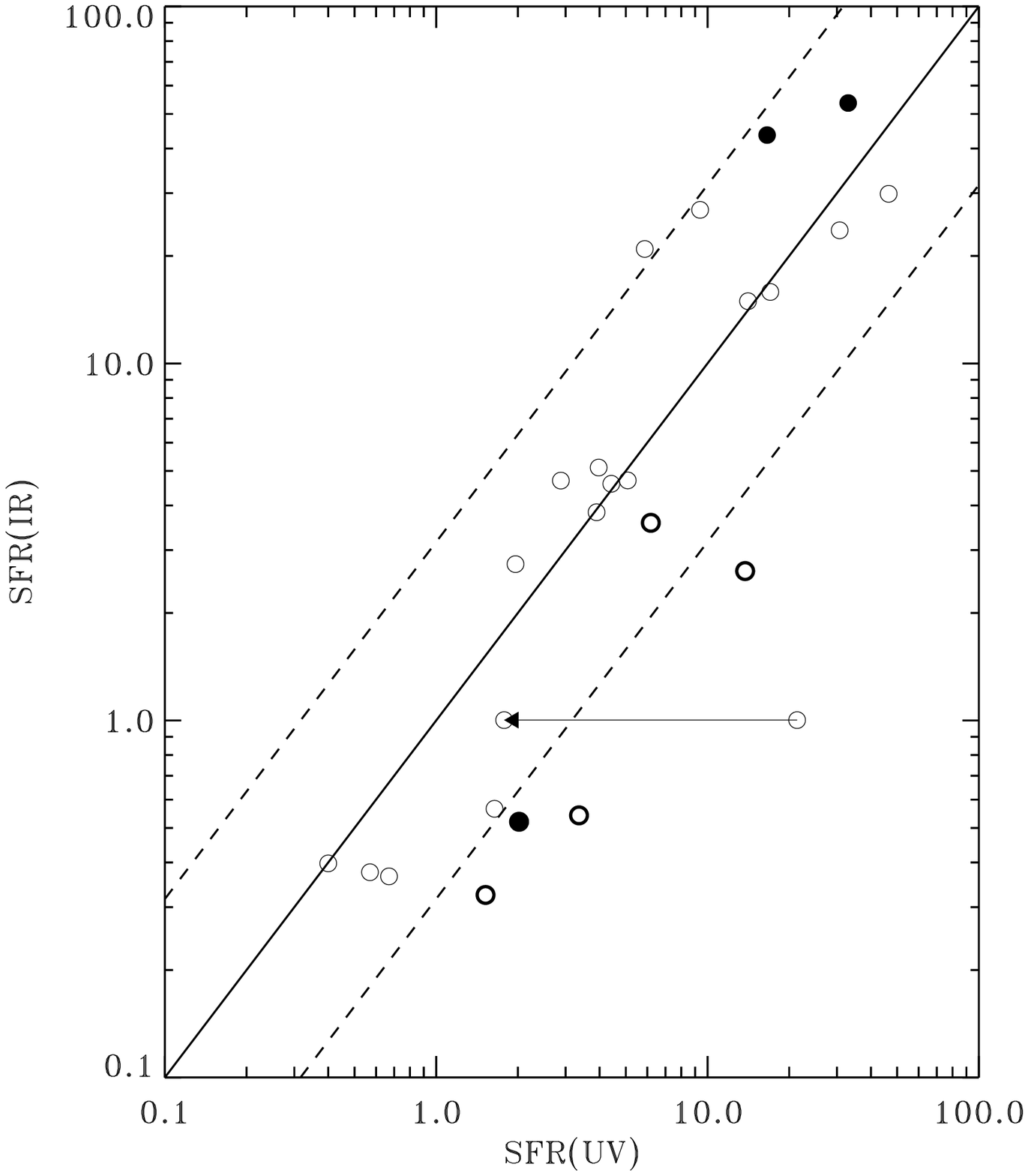}}
\end{picture}
\caption{\label{SFR:uv-IR} 
Comparison between the SFR(IR) and the SFR(UV) for the galaxies
detected by Spitzer. 
Filled circles are the galaxies with  $\Delta$XUV greater than 1.
Open circles are galaxies for which 
we extracted the 24 $\mu$m flux from the provided  maps. 
The solid line represents equal values and the dashed lines a 
deviation of 0.5 dex. 
The arrow connects the SFR(UV) calculated 
using F2000 with the SFR(UV) using  F2000$^*$ for CDFS270. 
}
\end{figure}

The SFR(IR) computed using equation 9 is given in 
Table~\ref{Tab:CDFS-Spitzer} and 
plotted against the SFR(UV) in Figure~\ref{SFR:uv-IR}.
In general we find a good correlation between the SFR(UV) and the 
SFR(IR) for most of the sample galaxies detected by Spitzer.  

CDFS088 and CDFS240, both with $\Delta$XUV greater 
than 1, were detected by Spitzer and have a SFR(IR) similar to the SFR(UV), 
confirming that the observed X-ray luminosity is probably dominated by 
nuclear activity while both UV and FIR are dominated by the stellar 
energy input.
Unfortunately CDFS017, CDFS304 and CDFS325 were
outside the Spitzer fields and therefore not observed.

There is one galaxy, CDFS270, that is quite far from the correlation 
showed in Figure~\ref{SFR:uv-IR}.
The arrow plotted in Figure~\ref{SFR:uv-IR} shows the result of 
applying the relation~\ref{eq:NewAv} to estimate the extinction 
in the UV range and to
recalculate the luminosity and the corresponding SFR(UV) as
explained in 6.1.

\begin{table*}
\begin{center}\caption{\label{Tab:CDFS-Spitzer}
IR data from Spitzer. The columns are: 
galaxy name, Spitzer index from the DR3, flux and error at 24 
$\mu$m, IR luminosity and estimated star formation rate.}
\begin{tabular}{lccccr}\hline
Name   & Spitzer Index & S$_{24}$    &     Error   & Log L $_{\rm IR}$   & SFR(IR) \\
       &               & (microJy)   &   (microJy) & (erg s$^{-1}$)  & (\Myear)\\
\hline
 CDFS065 & 279 &  1.06E+03 &  8.28E+00 &  44.52 & 15    \\
 CDFS073 & 784 &  1.00E+02 &  3.28E+00 &  44.02 &  4.7  \\
 CDFS078 & 000 &  81.0E+00 &  12.0E+00 &  43.06 &  0.52 \\
 CDFS088 & 127 &  7.74E+02 &  9.36E+00 &  45.08 & 54    \\
 CDFS129 &  75 &  3.31E+03 &  2.13E+01 &  44.67 & 21    \\
 CDFS132 & 000 &  28.0E+00 &  5.60E+00 &  42.86 &  0.32 \\
 CDFS149 & 850 &  1.40E+02 &  4.21E+00 &  42.92 &  0.38 \\
 CDFS152 & 000 &  40.0E+00 &  04.8E+00 &  43.08 &  0.54 \\
 CDFS158 & 289 &  2.25E+02 &  4.71E+00 &  44.06 &  5.1  \\
 CDFS167 & 000 &  21.0E+00 &  4.20E+00 &  43.77 &  2.6  \\
 CDFS185 & 000 &  31.0E+00 &  4.30E+00 &  43.90 &  3.6  \\
 CDFS189 & 189 &  7.59E+03 &  7.59E+02 &  44.01 &  4.6  \\
 CDFS192 & 178 &  4.66E+03 &  4.66E+02 &  43.93 &  3.8  \\
 CDFS196 & 304 &  1.40E+02 &  5.65E+00 &  44.55 & 16    \\
 CDFS207 & 159 &  1.93E+02 &  4.32E+00 &  42.95 &  0.40 \\
 CDFS213 & 548 &  7.09E+03 &  7.09E+02 &  43.81 &  2.9  \\
 CDFS236 & 309 &  8.35E+02 &  6.62E+00 &  44.78 & 27    \\
 CDFS238 & 376 &  5.43E+02 &  4.62E+00 &  44.02 &  4.7  \\ 
 CDFS240 & 581 &  3.74E+03 &  4.34E+01 &  44.99 & 44    \\
 CDFS265 & 656 &  9.69E+02 &  6.88E+00 &  44.82 & 30    \\
 CDFS267 & 549 &  2.99E+02 &  3.81E+00 &  42.91 &  0.37 \\
 CDFS270 & 105 &  5.59E+02 &  4.97E+00 &  43.35 &  1.0  \\
 CDFS291 & 335 &  3.51E+02 &  5.95E+00 &  43.10 &  0.57 \\
 CDFS292 & 487 &  7.00E+02 &  9.35E+00 &  44.72 & 24    \\
\hline
\end{tabular}
\end{center}
\end{table*}

\section{Conclusions}\label{Sec:Conclusions}

By comparing the  SFR(UV) with the SFR(X) in starforming
galaxies selected from the CDFS,  
we have confirmed the use of the X-ray luminosity as a reliable tracer of 
the current SFR even for galaxies with SFR as low as 1~\Myear.

%
The observed deviations can be explained by two different scenarios.
In one case we are detecting obscured AGN in which 
the nuclear activity is not affecting    
the observed UV-optical SED. In this case the SFR(X) is at least 
one order of magnitude higher than the SFR(UV).
For two of the sample galaxies (CDFS088 and CDFS240) 
where the $\Delta$XUV is greater than 1, 
the SFR(IR) is similar to 
the SFR(UV) pointing to the fact that the nuclear activity is
only affecting the output in the X-ray energy range.

We used empirical relations between UV anf IR fluxes
to set limits to the correction
factors applied  to the UV luminosities and the corresponding SFR(UV)
and found that in 
most of the objects with negative values of $\Delta$XUV the UV fluxes 
were overcorrected.

An extreme galaxy was found (CDFS213) 
for which the X-ray luminosity is very low compared to the expected one
based on SFR(UV).  
In this case we propose that the low X-ray luminosity is related to the delay 
between the peak of UV emission from massive stars  -- proportional to the 
SFR(UV) --
and the later onset of the X-ray emission related to the formation of the 
first 
HMXBs. Thus our proposal implies that in CDFS213 
we are probably witnessing a very young burst that is 
intense in UV but has not yet fully developed its HMXB population. 
This suggests that the age of the burst is probably shorter
than $\sim$20 Myrs. 


\section{Acknowledgments}  

We thank M\'onica Rodr\'\i guez, Divakara Mayya and Olga Vega for useful 
discussions. 
This research was partially supported by the French project 
PICS MEXIQUE \# 2174 for collaborative research.  
Computer facilities for DRG were kindly provided by 
{\it Dos-Inform\'atica}, Tenerife.
DRG, RT and ET acknowledge support by the Mexican
research council (CONACYT) under grants 49942 and  40018.
The GALEX data presented in this paper were obtained 
from the Multimission Archive at the Space Telescope Science Institute 
(MAST). STScI is operated by the Association of Universities for Research in 
Astronomy, Inc., under NASA contract NAS5-26555. 
Support for MAST for non-HST data is provided by the NASA Office of 
Space Science via grant NAG5-7584 and by other grants and contracts.

The authors are very grateful to an 
anonymous referee whose comments and suggestions largely improved the clarity
of this manuscript.

\bsp  
\label{lastpage}  
\end{document}